\documentclass[Preprint,12pt]{elsarticle}
\usepackage{newtxtext}

\usepackage{lineno}

\usepackage[bottom]{footmisc} 

\usepackage{geometry}
\usepackage{amsmath}
\usepackage{algorithm}
\usepackage{algpseudocode}
\usepackage{url}
\usepackage{ upgreek }
\usepackage{pdflscape}
\usepackage{caption}
\usepackage{setspace}
\usepackage{pifont}
\usepackage{amsmath}
\usepackage{bbding}
\usepackage{amsfonts} 
\usepackage{amssymb}
\usepackage{mathrsfs}

\usepackage{color}
\usepackage{ dsfont }
\usepackage[misc]{ifsym}

\usepackage{amsmath}      
\usepackage{graphicx}  
\usepackage{array}
\usepackage{verbatim}
\usepackage{titletoc}
\usepackage{color,xcolor}

\usepackage{units}
\usepackage{multirow}
\usepackage{tabularx}
\usepackage{booktabs}
\usepackage{bbm}
\usepackage{amsmath}      
\usepackage{bm} 
\usepackage{titlesec}
\usepackage{algorithm}
\usepackage{algpseudocode}
\titleformat{\paragraph}
  {\normalfont\normalsize\itshape}{\theparagraph}{1em}{}
\titlespacing*{\paragraph}
  {0pt}{1.5ex plus 0.5ex minus .2ex}{0.5ex plus .2ex}  

\usepackage{amsmath}

\usepackage{xcolor}
\definecolor{DeBlue}{RGB}{0,0,0}
\definecolor{DeGreen}{RGB}{84,130,53}
\definecolor{DeBlack}{RGB}{0,0,0}

\usepackage{natbib}
\usepackage{fancyhdr}

\begin{document}

\title{EV-STLLM: Electric vehicle charging forecasting based on spatio-temporal large language models with multi-frequency and multi-scale information fusion} 

\begin{frontmatter}

\author[inst1,inst2]{Hang Fan}
\affiliation[inst1]{organization={School of Economic and Management},
            addressline={North China Electric Power University}, 
            city={Changping District, Beijing, 102206},
            country={China}}

\affiliation[inst2]{organization={Beijing Key Laboratory of Renewable Energy and Low-carbon Development},
            addressline={Changping District}, 
            city={Beijing, 102206},
            country={China}}

\affiliation[inst3]{organization={School of Energy Power and Machinery Engineering},
            addressline={North China Electric Power University}, 
            city={Changping District, Beijing, 102206},
            country={China}}

\affiliation[inst4]{organization={College of Computing and Data Science},
             addressline={Nanyang Technological University}, 
             city={50 Nanyang Avenue, 639798},
             country={Singapore}}

\affiliation[inst5]{organization={School of Electrical and Electronic Engineering},
             addressline={Nanyang Technological University}, 
             city={50 Nanyang Avenue, 639798},
             country={Singapore}}
             
\author[inst3]{Yunze Chai}
\author[inst4]{Chenxi Liu}
\author[inst5]{Weican Liu\corref{cor1}}\footnotesize
\cortext[cor1]{W. Liu is the corresponding author (Email: weican001@e.ntu.edu.sg). This work is supported by the Basic Research Operations of Central Universities (2024MS027).}

\author[inst1]{Zuhan Zhang} 
\author[inst1]{Wencai Run}
\author[inst1,inst2]{Dunnan Liu} 
\normalsize

\begin{abstract}
With the proliferation of electric vehicles (EVs), accurate charging demand and station occupancy forecasting are critical for optimizing urban energy and the profit of EVs aggregator. Existing approaches in this field usually struggle to capture the complex spatio-temporal dependencies in EV charging behaviors, and their limited model parameters hinder their ability to learn complex data distribution representations from large datasets. To this end, we propose a novel EV spatio-temporal large language model (EV-STLLM) for accurate prediction. Our proposed framework is divided into two modules. In the data processing module, we utilize variational mode decomposition (VMD) for data denoising, and improved complete ensemble empirical mode decomposition with adaptive noise (ICEEMDAN) for data multi-frequency decomposition. Fuzzy information granulation (FIG) for extracting multi-scale information. Additionally, ReliefF is used for feature selection to mitigate redundancy. In the forecasting module, the EV-STLLM is used to directly achieve EV charging and occupancy forecasting. Firstly, we fully capture the intrinsic spatio-temporal characteristics of the data by integrating adjacency matrices derived from the regional stations network and spatio-temporal-frequency embedding information. Then, the partially frozen graph attention (PFGA) module is utilized to maintain the sequential feature modeling capabilities of the pre-trained large model while incorporating EV domain knowledge. Extensive experiments using real-world data from Shenzhen, China, demonstrate that our proposed framework can achieve superior accuracy and robustness compared to the state-of-the-art benchmarks.
\end{abstract} 


\end{frontmatter}
\setcounter{page}{1}

\section{Introduction}

\subsection{Background and motivation}
The rapid proliferation of electric vehicles (EVs) has led to a substantial increase in the demand for efficient charging infrastructure\cite{GUO2015420, fan2025incentive}. However, the stochastic and complex spatio-temporal dependent charging behaviors pose significant challenges for the management of the charging infrastructure\cite{LI2025125933, WU2023116619}. Accurate forecasting of EV charging demand including the charging volume and occupancy rate is crucial to address the challenge. It can benefit not only power grid operators but also the charging service operators\cite{SHANG2025125460}. For grid operators, predicting future charging volumes helps to find potential high-load events, enabling proactive network regulation to ensure grid stability. For charging service operators, occupancy forecasts can guide dynamic pricing strategies, such as increasing fees during peak demand to manage user flow and decreasing them during off-peak times to attract more customers, thereby optimizing revenue and operational efficiency. Consequently, developing robust and generalizable forecasting models for EV charging has become essential for the strategic scheduling and allocation of public charging resources.

\vspace{0.25\baselineskip}

Despite advances in EV charging demand prediction, these methods often face critical limitations. Existing models frequently struggle to capture the complex, multi-dimensional spatio-temporal dependencies in charging load \cite{WANG2024130305}. They find it difficult to simultaneously model the combined effects of different cycles (e.g., daily vs. weekly) and multiple frequencies (e.g., short-term fluctuations vs. long-term trends). More importantly, the limited parameter scale of these models constrains their ability to learn complex data distributions and deep representations from large-scale datasets \cite{WANG2025134884}. In contrast, Large Language Models (LLMs) are characterized by their large parameter scale and sophisticated architectures, granting them superior representation learning capabilities essential for deeply understanding and analyzing complex data patterns. To bridge the gap and apply the power of LLMs to EV prediction, we introduce a partially frozen graph attention module. This design leverages the remarkable cross-domain knowledge transfer inherent in LLMs. Specifically, the partially frozen mechanism retains and fine-tunes pretrained LLM knowledge, making the model adaptable to specific EV forecasting tasks. Complementarily, the graph attention mechanism is employed to integrate domain-specific knowledge, such as the spatial structure of the charging station network, thereby enabling the model to align heterogeneous features within a unified semantic space. By combining the general representational capacity of LLMs with specialized domain graph attention, our proposed framework effectively mitigates the representation bottlenecks prevalent in existing models, resulting in enhanced prediction accuracy and generalizability.

\subsection{Related work}
Driven by the rising adoption of EVs, recent years have seen substantial research growth in EV charging forecasting. These studies can be broadly categorized into four types of models: physical models, statistical models, artificial intelligence (AI)-based models, and hybrid models \cite{Jia2023ReviewOptimizationForecasting, xing2021modelling, Li2022ReviewLoadForecastingMethods, Liu2025}. 
Physical models mainly rely on the physical representation of the movement process of vehicles in the traffic network and their energy consumption. For example, Tang \textit{et al.} investigated the relationship between EV power, vehicle velocity, acceleration, and road grade, proposed an analytical model for charging power estimation, which effectively estimates the vehicle's energy consumption based on physical process\cite{WU201552}. 
While these approaches offer utility, they are generally more suitable for specific conditions or scenarios. This is because they rely solely on the physical parameters of the battery, thereby overlooking the complex influence of human behavior and activities on EV charging load prediction.
In contrast, statistical models have been widely adopted for EV charging forecasting and have demonstrated promising accuracy. 
These models usually rely on rigorous mathematical theory derivation, considering key variables such as charging start time, State of Charge (SOC), and the total number of charging EVs. A significant advantage of these models is their relatively small parameter count, which generally allows for effective training with less historical data. \cite{WeicanEnhancing}. For example, Md Shariful Islam \textit{et al.} extracts the charging characteristics from statistical data features from previous data of EVCL, and afterwards uses these features to simulate the SOC distribution of EVs and finally predicts EVCL by SOC before day, which can reduce the number of decision variables significantly, and require less computational time and memory accordingly \cite{8061012}. Yan \textit{et al.} proposed an EV charging forecasting method that establishes a stochastic model for PEV charging and derives a probabilistic description of electricity needs for one charging station at any hour. This approach utilized actual survey data from the National Household Travel Survey (NHTS) to build statistical models for each key variable, achieving superior prediction accuracy in estimating electricity consumption for a single charging station \cite{2017Statistical}.
With the continuous advancement of deep learning technologies, AI-based models have gradually emerged as mainstream approaches in EV charging forecasting, owing to their powerful capabilities in modeling the nonlinear characteristics of EV charging data \cite{10690538}. Several fundamental deep learning models have demonstrated their effectiveness in capturing complex EV charging patterns in previous studies, including support vector machines(SVM) \cite{SVM}, artificial neural networks (ANN) \cite{ANN}, and long and short-term memory (LSTM) \cite{LSTM}. For example, Sun \textit{et al.} used SVM to preprocess multi-source input data and effectively correct for anomalies, thereby improving the accuracy of the charging load prediction model \cite{SVM2}.
However, AI-based models use only a single model, which cannot show good generalization in different data sets. Therefore, hybrid models have begun to attract the attention of researchers, such as the hybrid CNN-LSTM-Attention deep learning model \cite{2024HybridCNN} and the ARIMA-LSTM model \cite{2025HybridARIMA}. For example, Tian \textit{et al.} proposed a combination of Temporal Convolutional Network (TCN) and LSTM, called TCN-LSTM, which has notably improved for the specific task \cite{TIAN2025125174}. Ge \textit{et al.} proposed the SeqGAN-LSTM framework when data is missing, which significantly improves the accuracy and robustness of charging load forecasting \cite{10667370}. Yin \textit{et al.} proposed a model based on federated learning and variational mode decomposition-long short-term memory neural network, effectively improving the accuracy of short-term electric vehicle charging load forecasting while ensuring user privacy security \cite{YIN2025134559}.

\vspace{0.25\baselineskip}
Despite significant advancements in forecasting EV charging loads, practical application of current methodologies faces several critical challenges.
First, due to the limitation of the number of parameters, it is difficult for existing models to fully learn the full data distribution information of EV charging datasets. In addition, these small single models can only perform well in a few datasets, therefore, when generalizing to other different EV charging datasets, they often cannot guarantee accuracy and have difficulty in showing good general characteristics \cite{he2016deep}. 
Second, they have difficulty understanding the relationships between heterogeneous data \cite{2024PAG}, such as time series, geographical information, and socio-economic factors. This hinders comprehensive prediction of charging demand and charging-related management.
To address these shortcomings, this study advocates for the application of LLMs. LLMs possess a unique capability to reframe the EV charging demand prediction task as a natural language processing (NLP) problem, thereby leveraging their robust semantic understanding to process and interpret varied data types (e.g., numerical time series, categorical events, textual descriptions) \cite{Li2025UrbanEV}. Furthermore, the extensive pre-training on vast text corpora endows LLMs with exceptional generalization abilities, allowing them to discern subtle patterns and dependencies in data, even when faced with novel scenarios \cite{liuSTLLM}. This inherent capacity for feature learning significantly reduces the need for laborious and often domain-specific feature engineering common in conventional forecasting paradigms. 
More recently, inspired by the success of incorporating spatial information with temporal patterns in traffic prediction, spatio-temporal EV charging demand prediction has emerged as an attractive research topic in the literature. Representative examples include HSTGCN-EV \cite{wang2023predicting} and ChatEV \cite{qu2024chatev}: The former one incorporated two heterogeneous graphs (i.e., a demand-based graph and a geographic graph) to improve predictive precision, while the latter one unified spatial and temporal factors within natural language and harnessed LLMs for regional EV charging prediction. 
While recent LLM-based approaches have shown promise, they often rely on prompt engineering to fuse heterogeneous data. ChatEV, which converts diverse inputs like weather and historical sequences into natural language, tends to be superficial. It particularly falls short in two critical areas: it fails to explicitly leverage the physical network topology of charging stations, and it does not perform deep signal processing to uncover the complex, multi-frequency dynamics within the time series data itself.
To overcome these limitations, we propose the EV-STLLM, which introduces two targeted innovations. Firstly, to address the superficial temporal analysis, our data preprocessing module employs the VMD-ICEEMDAN technique. This method decomposes the original time series into distinct high, medium, and low-frequency components, enabling the model to discern between short-term fluctuations, periodic cycles, and long-term trends. Secondly, to counteract the lack of structural awareness, our Partially Frozen PFGA module explicitly integrates the network topology by using the station adjacency matrix as a mask within the attention mechanism. This design compels the model to learn spatially-aware representations, ensuring a deeper and more accurate understanding of spatio-temporal dependencies.
Furthermore, existing loss functions for EV charging forecasting are typically based on the time dominated paradigm, 
which can struggle to fully capture complex characteristics of EV charging data in the frequency domain. What's more, the prediction paradigm based on direct prediction usually ignores the autocorrelation of label sequences. 
Therefore, we have proposed a customized loss function based on the time and frequency domains, which not only preserves the advantages of the direct prediction paradigm but also effectively suppresses the expression of autocorrelation in the label sequence. \textbf{Table 1} contrasts our work with the literature.

\begin{table}[h]
\small
\begin{spacing}{1}
\caption{\textbf{Contrasting our work with the literature}}
\resizebox{\textwidth}{!}{%
\setlength{\tabcolsep}{2mm}{
\begin{tabular}{ccccccc}
\hline
\multicolumn{2}{c}{\multirow{3}{*}{Reference}}                                                                                                                        & \multirow{3}{*}{\begin{tabular}[c]{@{}c@{}}Feature \\ Engineering\end{tabular}} & \multirow{3}{*}{\begin{tabular}[c]{@{}c@{}}Data \\ Preprocessing\end{tabular}} & \multirow{3}{*}{\begin{tabular}[c]{@{}c@{}}Multi-information\\ Fusion\end{tabular}} & \multirow{3}{*}{\begin{tabular}[c]{@{}c@{}}Spacial-Temproal\\ Embedding \end{tabular}} & \multirow{3}{*}{\begin{tabular}[c]{@{}c@{}}Based on\\ LLMs\end{tabular}} \\
\multicolumn{2}{c}{}                                                                                                                                                  &                                                                                 &                                                                                                                                                     &                                                                                 &                                                                              &                                                                             \\
\multicolumn{2}{c}{}                                                                                                                                                                                                                                   &                                                                                &                                                                      &                                                                                 &                                                                              &                                                                             \\ \hline
\textit{\begin{tabular}[c]{@{}c@{}} Physical \& \\ Statistical \\ Model\end{tabular}} & \cite{WU201552, 8061012, 2017Statistical} & \ding{55}                                                          & \ding{55}                                                         & \ding{55}                                                                                                     & \ding{55}                                                       & \ding{55}                                                   \\
\multirow{5}{*}{\textit{AI Based}}                               & \cite{10690538, SVM, ANN, LSTM, SVM2} & \checkmark                                                      & \checkmark                                                                                                      & \ding{55}                                                       & \ding{55}                                                       & \ding{55}                                                    \\
 & \cite{2024HybridCNN, 2025HybridARIMA, TIAN2025125174, 10667370, YIN2025134559,2024PAG}                                          &  \checkmark                                                   & \checkmark                                                                  & \ding{55} & \ding{55}                                                       & \ding{55}                                                       \\
 & \cite{qu2024chatev, Li2025UrbanEV}                                          & \checkmark                                                    & \checkmark                                                      & \ding{55}                & \ding{55}                                                       & \checkmark                                                      \\
 & \cite{liuSTLLM, ST-LLM+}                                        & \checkmark                                                      & \checkmark                                                          & \ding{55}                                                                                                 & \checkmark                                                       & \checkmark                                                      \\
 & \textbf{Our Work (EV-STLLM)}                                                         & \textbf{\checkmark}                                              & \textbf{\checkmark}                                                                              & \textbf{\checkmark}                                              & \textbf{\checkmark}                                           & \textbf{\checkmark}                                          \\ \hline
\end{tabular}}
}
\end{spacing}
\end{table}

\subsection{Contributions}

In this paper, we propose EV-STLLM, a novel framework for EV charging forecasting that leverages the power of spatio-temporal large language models combined with advanced data processing techniques. Our approach is designed to overcome the limitations of existing models by effectively capturing the complex, multi-faceted dependencies inherent in EV charging data. The EV-STLLM consists of two main stages: an effective data processing module that extracts and fuses multi-frequency and multi-scale information, and a customized spatio-temporal LLM that accurately predicts future charging demand and station occupancy. By integrating graph-based attention and efficient fine-tuning strategies like QLoRA, our model can adapt the extensive knowledge of pre-trained LLMs to the specific characteristics of the EV charging domain. 
\vspace{0.25\baselineskip}

\textbf{The main contributions of this paper can be summarized as follows:} 
\begin{itemize}
 
\item 
\emph{\textbf{A novel EV charging forecasting framework via EV-STLLM for grid management}}: We propose a novel EV charging forecasting framework, EV-STLLM, designed for enhanced grid management through accurate and robust predictions. Our proposed framework is mainly divided into two parts. First, the data preprocessing module is dedicated to extracting rich information from raw data by fusing multi-frequency and multi-scale features. Second, we proposed a customized spatio-temporal large language model, which is specifically tailored to interpret the complex spatio-temporal dependencies within the preprocessed data, leading to a holistic and deep understanding of EV charging patterns for the final prediction. Using a dataset from Shenzhen, China, we verified the effectiveness of our proposed prediction framework. 

\item 
\emph{\textbf{The multi-frequency and multi-scale information fusion strategy for data preprocessing}}: 
To fully extract the different frequency representation information and different time scale information of the data, we proposed this learning strategy. In multi-frequency extraction, we employ a hybrid VMD-ICEEMDAN technique to decompose and reconstruct the signal into distinct high, mid, and low-frequency components, capturing different underlying volatilities. In multi-scale extraction, we utilized FIG to generate data representations at different temporal granularities (e.g., daily and weekly), revealing both short-term fluctuations and long-term trends. Finally, to mitigate feature redundancy, we utilized the ReliefF algorithm to identify and retain the most impactful additional features.

\item  
\emph{\textbf{The customized spatio-temporal LLM for accurate EV charging forecasting}}: We proposed a customized spatio-temporal LLM based on the pre-trained LLM, which can be divided into two parts. First, the proposed PFGA module strategically modifies the LLM architecture. In freezing layers, the model preserves general sequence modeling capabilities. In unfreezing layers, the model incorporates the graph-based attention mechanism, which effectively learns domain-specific spatial dependencies from the charging station network. Second, to ensure computational feasibility, we integrate the Quantized Low-Rank Adaptation (QLoRA) fine-tuning strategy, which dramatically reduces memory overhead and trainable parameters by using 4-bit quantization, enabling efficient adaptation of the LLM without sacrificing predictive performance.

\end{itemize}
The rest of the paper proceeds as follows: Section 2 introduces the problem and the EV-STLLM's overall framework. Section 3 details the data processing module, the customized EV-STLLM for EV charging, and the tailored loss function for time-frequency domain feature learning. Section 4 outlines the data selection, parameter design, evaluation metrics, and operational environments. Section 5 then showcases the proposed EV-STLLM's performance through four experiments. Finally, Section 6 concludes our work and explores avenues for future research.

\section{Problem Formulation and Framework Design} 

In this section, we provide a detailed introduction to the problem formulation that our research addresses, along with a thorough explanation of the framework design for our proposed model. \textbf{Table 2} lists the key notations.

\subsection{Problem formulation} 

In this work, our objective is to predict the EV charging demand (volume) and EV charging occupancy rate, measured in $\%$ and $kWh$, respectively.

\textit{\textbf{Definition 1.} EV charging Data}: We denote the EV charging data as $\mathbf{X} \in \mathbb{R}^{T \times N \times C}$, where $T$ is the number of time steps, $N$ is the number of spatial stations, and $C$ is the EV charging feature. For example, $C = 1$ represents the EV charging demand or occupancy rate.

\vspace{0.25\baselineskip}
\textit{\textbf{Definition 2.} EV charging network}: We formulate EV charging network as a graph $G = (V, E, \mathbf{A})$, where $V$ denotes a set of $N$ nodes, each representing a spatial zone. The set of edges is indicated by $E \subseteq V \times V$, and $\mathbf{A} \in \mathbb{R}^{N \times N}$ encodes the adjacency relationships reflecting spatial proximities among the zones.

\vspace{0.25\baselineskip}
\textit{\textbf{Definition 3.} EV charging Forecasting}: Given the historical EV charging data of $P$ time steps $\mathbf{X}_P = \{\mathbf{x}_{t-P+1}, \mathbf{x}_{t-P+2}, \ldots, \mathbf{x}_t\} \in \mathbb{R}^{P \times N \times C}$ and EV~charging network $G$, the objective is to learn a function $f(\cdot)$ with parameter $\theta$ to predict EV charging data of the following $S$ time steps $\mathbf{Y}_S = \{\mathbf{y}_{t+1}, \mathbf{y}_{t+2}, \ldots, \mathbf{y}_{t+S}\} \in \mathbb{R}^{S \times N \times C}$.
\begin{equation}
\left[\mathbf{x}_{t-P+1}, \mathbf{x}_{t-P+2}, \ldots, \mathbf{x}_t, G\right] \xrightarrow[\theta]{f(\cdot)}\left[\mathbf{y}_{t+1}, \mathbf{y}_{t+2}, \ldots, \mathbf{y}_{t+S}\right].
\end{equation}

\begin{table}[!t]
    \caption{Summary of notations}
    \label{tab:notations}
    \setlength{\tabcolsep}{1.9mm}{
    \begin{tabular}{llll}
        \toprule
        \textbf{Notation} & \textbf{Description} & \textbf{Notation} & \textbf{Description} \\
        \midrule
        $\mathbf{X}$ & EV charging Data & $G$ & EV Graph \\
        $N$ & Number of spatial locations in $\mathbf{X}$ & $V$ & Nodes or spatial locations of $G$ \\
        $T$ & Number of time steps in $\mathbf{X}$ & $E$ & Edges of $G$ \\
        $C$ & Number of features in $\mathbf{X}$ & $\mathbf{A}$ & Proximity-based adjacency matrix of $G$ \\
        $P$ & Number of historical time steps & $\mathbf{E}_P$ & Token embedding of historical $P$ time steps \\
        $S$ & Number of future time steps & $\mathbf{E}_T$ & Temporal embedding \\
        $\mathbf{X}_P$ & Historical EV charging Data & $\mathbf{E}_S$ & Spatial embedding \\
        $\mathbf{Y}_S$ & Future EV charging Data & $\mathbf{H}^L$ & Final output of the PFGA LLMs Module \\
        \bottomrule
    \end{tabular}}
\end{table}

\subsection{Framework design of EV-STLLM} 
\begin{figure}[!t]
    \centering
    \captionsetup{labelfont=bf}
    \includegraphics[width=0.82\linewidth]{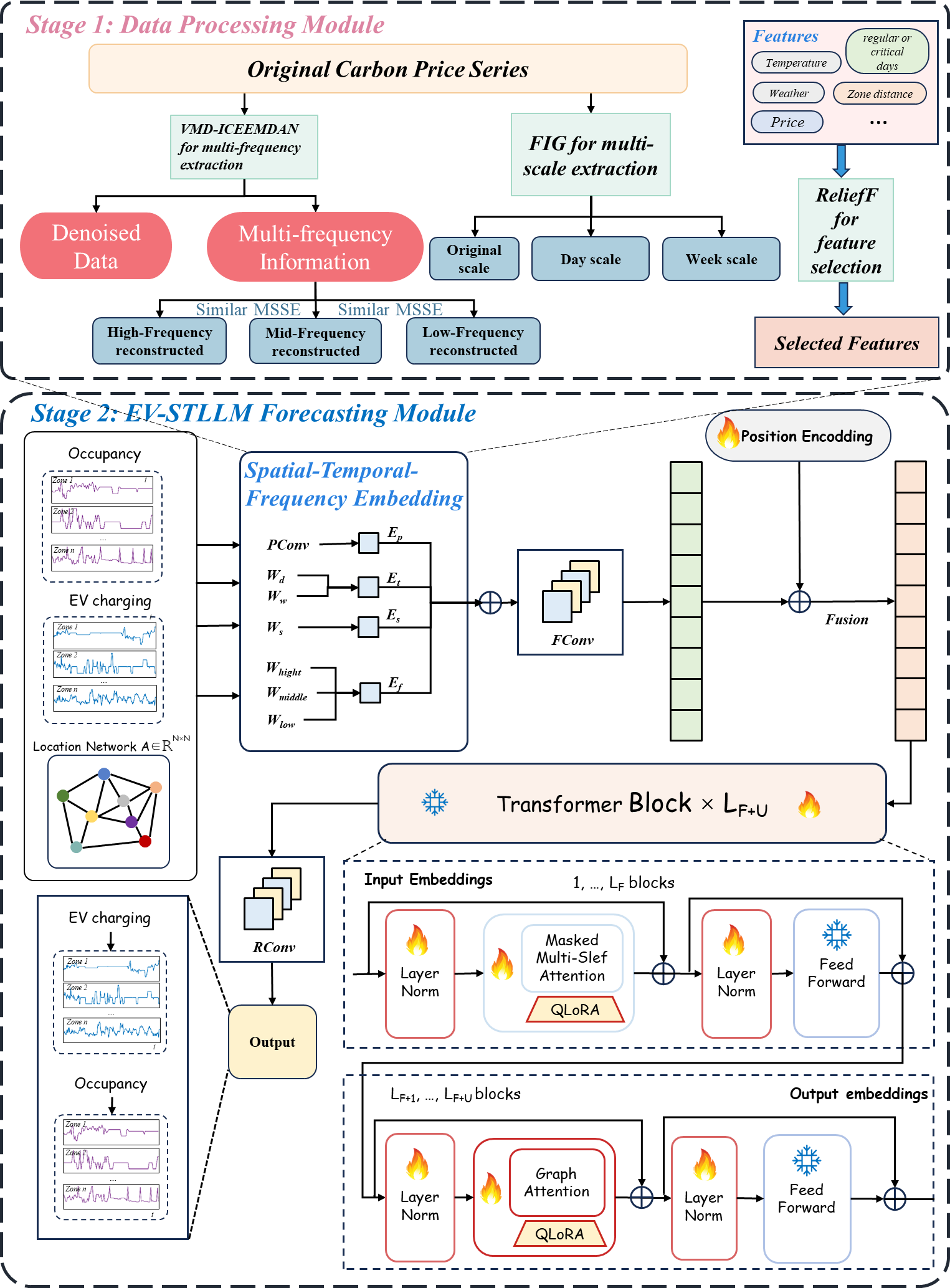}
    \vspace{-0.1\baselineskip}
    \caption{\textbf{Flow chart of EV-STLLM}}
\end{figure} 

Our proposed EV-STLLM is primarily divided into two modules: the data preprocessing module, the EV-STLLM forecasting module. In the data preprocessing module, raw EV charging data is initially denoised using VMD by removing the highest frequency white noise component. The denoised data is then analyzed for complexity with Multi-Scale Sample Entropy (MSSE), and the most complex component undergoes further decomposition using the ICEEMDAN algorithm. Subsequent sub-components with similar MSSE values are grouped into high, mid, and low-frequency time series, capturing the data's dynamic temporal structures. Feature engineering is enhanced with ReliefF, reducing redundancy and incorporating domain knowledge for model training. 
In the EV-STLLM forecasting module, data passes through a Spatio-Temporal-Frequency Embedding layer to create embeddings, which are fused in a convolution layer for advanced representation. This output, along with a charging station adjacency matrix, is processed by a spatio-temporal LLM with PFGA, consisting of transformer blocks employing a Graph Attention mechanism to leverage spatial relationships. The QLoRA method optimizes the training of unfrozen layers, and the final output is refined through a regression convolution layer for EV charging and occupancy predictions. The process concludes with a time-frequency domain feature learning loss function to enhance prediction accuracy. \textbf{Figure 1} demonstrates the flow chart of our EV-STLLM.

\section{Methodology} 

In this section, we first introduce the data preprocessing module, the EV-STLLM forecasting module, and the customized time-frequency domain feature learning loss function.

\subsection{Data preprocessing}
The inherent complexity and non-linear patterns of EV charging data make it challenging for standard neural networks to extract meaningful insights. Specifically, existing methods fail to fully exploit the multi-frequency and multi-scale information inherent in the data. To this end, we propose a dedicated data processing module as depicted in \textbf{Figure 2}. 
Initially, it separates the original signal into Intrinsic Mode Functions (IMFs) using VMD, which mitigates signal processing issues such as modal aliasing and boundary effects. The high-frequency noise is removed to produce a denoised signal. The next step involves ICEEMDAN, which refines the extraction of IMF by adding adaptive noise, enhancing stability and precision.
Subsequently, the MSSE method is employed to quantify the complexity of the time series across multiple time scales. This process involves constructing new series representations at different scales with FIG, a technique that allows for noise suppression and the retention of essential data characteristics by using fuzzy sets for information extraction at varying granularities.
To improve model forecasting accuracy, the ReliefF algorithm is used to select the most predictive features, effectively handling multi-class data and mitigating feature redundancy issues. This algorithm evaluates feature interactions by analyzing the values of neighboring samples and updates feature weights based on nearest neighbors from both similar and different classes. The combined use of these techniques aims to enhance the predictive accuracy and robustness of models in handling the dynamic features of EV charging data.

\begin{figure}[!b]
\centering
\captionsetup{labelfont=bf}
\includegraphics[width=1.0\linewidth]{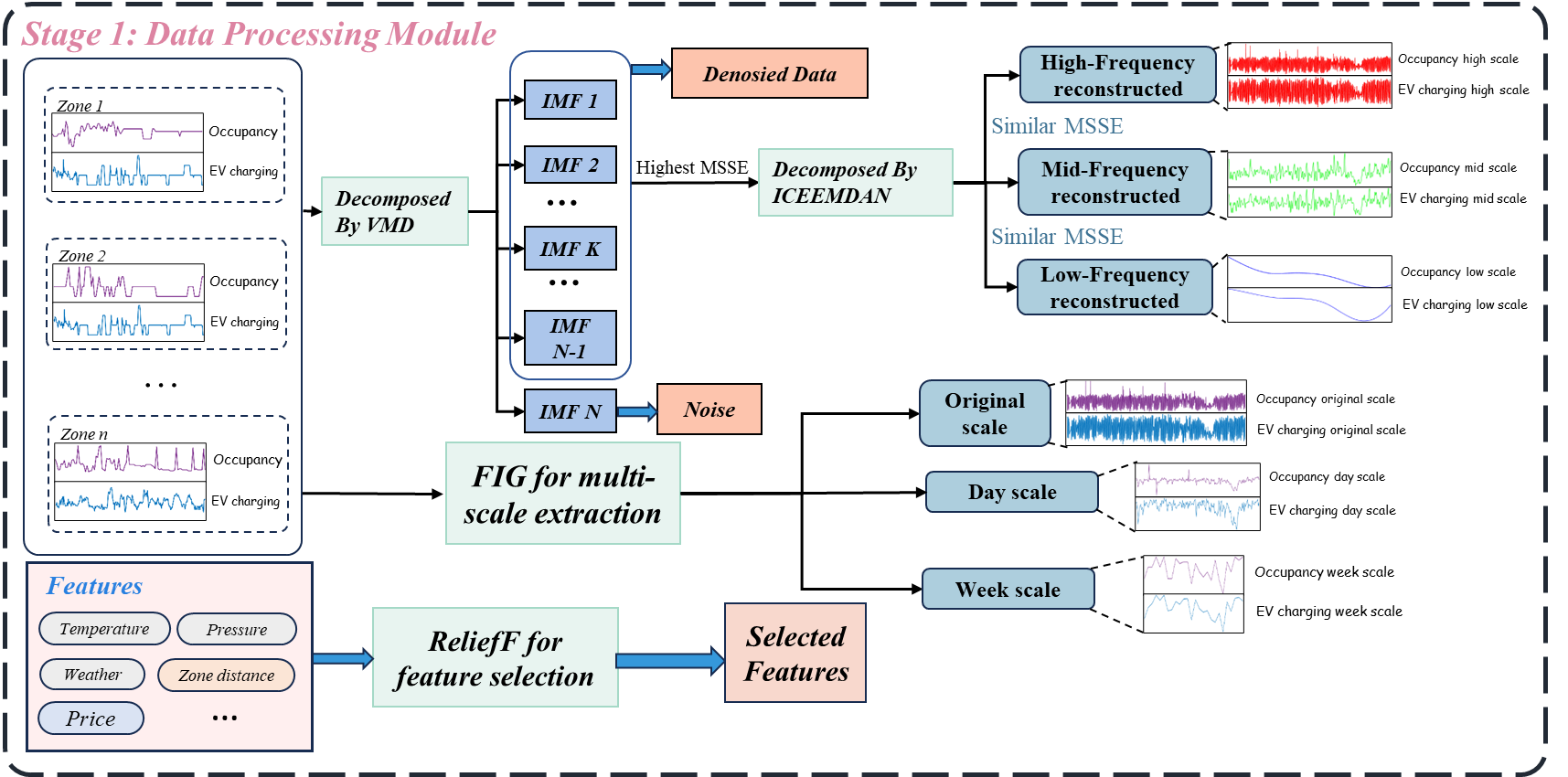}
\vspace{-0.1\baselineskip}
\caption{\textbf{Flow chart of the data preprocessing module}}
\end{figure}

\subsubsection{VMD-ICEEMDAN For Multi-frequency Extraction}
The multi-frequency extraction process is executed in three sequential steps: first, a primary decomposition and denoising of the raw signal using VMD; second, a targeted secondary decomposition of the most complex resulting component using ICEEMDAN, guided by MSSE; and third, a final recombination of all components into high, mid, and low-frequency series based on their MSSE similarity.

\vspace{0.25\baselineskip}
The initial step separates the original signal into a set of IMFs. Based on the assumption that high-frequency oscillations represent system noise, the IMF with the highest frequency is isolated and removed. The remaining IMFs (IMF 1 to IMF N-1) are then recombined to produce a clean, denoised signal. 
To deconstruct the signal into its constituent frequencies, we first employ VMD. As a technique based on solving a constrained variational problem, VMD is particularly adept at mitigating common signal processing issues like modal aliasing and boundary effects \cite{tian2024cnns}. It also offers the flexibility to define the number of decomposition layers and maintains strong robustness against noise. The core of VMD is to find a set of modes $\left\{u_k\right\}$ and their respective center frequencies $\left\{w_k\right\}$ that, in aggregate, reconstruct the original signal f. This is formulated as the following optimization problem:

\begin{equation}
\begin{aligned}
\min _{\left\{u_k\right\},\left\{w_k\right\}} & \left\{\sum_{k=1}^K\left\|\partial_t\left[\left(\delta(t)+\frac{j}{\pi t}\right) \times u_k(t)\right] e^{-j w_k t}\right\|_2^2\right\}, \\
\text { s.t. } & \sum_{k=1}^K u_k=f,
\end{aligned}
\end{equation}
where $\delta(t)$ is the Dirac function, $\partial_t$ is the partial time derivative, and $K$ is the total number of modes. To solve this constrained problem, it is converted into an unconstrained one using an Augmented Lagrangian function, which can be optimized via the Alternating Direction Method of Multipliers (ADMM):

\begin{equation}
\begin{aligned}
L\left(\left\{u_k\right\},\left\{w_k\right\}, \lambda\right)=\alpha & \sum_{k=1}^K\left\|\partial_t\left[(\delta(t)+j / \pi t) \times u_k(t)\right] e^{-j w_k t}\right\|_2^2 \\
& +\left\|f(t)-\sum_{k=1}^K u_k(t)\right\|_2^2+\left\langle\lambda(t), f(t)-\sum_{k=1}^K u(t)\right\rangle.
\end{aligned}
\end{equation}

\vspace{0.25\baselineskip}

Next, for a more refined decomposition, we utilize ICEEMDAN \cite{sun2024multivariate, al-musaylh2018two}. This method builds upon the foundation of Empirical Mode Decomposition (EMD) \cite{xu2024enhanced} by introducing adaptive noise to enhance the stability and precision of the signal separation process. The procedure involves two main stages:
Noise Addition and EMD Application. The process begins by adding independent white noise of EV charging data to the signal in each iteration. The standard EMD algorithm is then applied to this noise-assisted signal, $x_n(t)$, to extract a set of Intrinsic Mode Functions (IMFs):

\begin{equation}
x_n(t)=\sum_{i=1}^{M_n} \mathrm{IMF}_{i, n}(t)+r_{M_n, n}(t),
\end{equation}
where ${IMF}_{i, n}(t)$ is the $i$-th IMF derived from the noisy signal $x_n(t)$, and $r_{M_n, n}(t)$ is the final residual. After performing multiple iterations with different noise realizations, the final IMFs are determined by averaging the results. The original signal is then reconstructed as follows:

\begin{equation}
x(t)=\sum_{i=1}^K\left(\frac{1}{N} \sum_{n=1}^N \operatorname{IMF}_{i, n}(t)\right)+r_K(t),
\end{equation}
where $K$ represents the total count of IMFs extracted, $N$ is the number of noise-addition iterations, and $r_K(t)$ is the final residual component of the signal.

\vspace{0.25\baselineskip}

Finally, to guide the recombination of these decomposed components, we use MSSE. This metric quantifies the complexity of a time series by extending the concept of Sample Entropy (SampEn) across multiple time scales, offering a richer view of the signal's dynamic properties. For a given scale factor $\tau$, a coarse-grained time series is first generated:

\begin{equation}
y_j^{(\tau)}=\frac{1}{\tau} \sum_{i=(j-1) \tau+1}^{j \tau} x_i, \quad j=1,2, \ldots,\left\lfloor\frac{N}{\tau}\right\rfloor,
\end{equation}
where $y_j^{(\tau)}$ is the newly constructed series at scale $\tau$. The sample entropy, SampEn($m$, $r$, $N_\tau$), is then computed for each of these series. By plotting the resulting entropy values against their corresponding scale factors, an MSSE curve is produced, which illustrates how the complexity of the time series evolves across different scales.

\subsubsection{FIG for multi-scale extraction}

EV charging time series are often composed of dynamic features operating at different temporal granularities, including both short-term fluctuations and long-term cyclical trends. A model's ability to distinguish and leverage these different patterns is crucial for enhancing its predictive accuracy and robustness. To this end, we employ FIG \cite{cao2023hybrid} as a scale-transformation technique. This method allows us to construct new feature representations at coarser time scales, such as weekly and monthly views, to supplement the original denoised data. The core of FIG is its use of a fuzzy membership function, defined as:

\begin{equation}
A(x, a, m, b)= \begin{cases}0, & x<a \\ \frac{x-a}{m-a}, & a \leq x \leq m \\ \frac{b-x}{b-m}, & m<x \leq b \\ 0, & x>b\end{cases},
\end{equation}

The function defines a classic triangular fuzzy set using $a$ and $b$ as endpoints and $m$ as the peak. When a window size is specified, this function effectively granulates the time series, enabling information extraction at different scales (e.g., daily or weekly). Unlike conventional aggregation methods like simple averaging, a key advantage of FIG is its inherent noise suppression capability. This stems from the overlapping nature of fuzzy sets, which naturally smooths out random outliers and noise, thereby retaining essential data characteristics more effectively while filtering out unwanted interference.

\vspace{0.25\baselineskip}

\subsubsection{ReliefF for feature selection}
To enhance the forecasting accuracy of our model, it is crucial to identify and select the most predictive features from the available data, thereby mitigating the negative impact of feature redundancy. To this end, we employ the ReliefF algorithm to screen all external features and isolate the most valuable components. These external features include not only typical numerical EV charging data but also a critical indicator variable derived from domain knowledge: holidays are set to 1, while ordinary days are set to 0.
ReliefF is particularly effective due to its exceptional capability in handling feature interactions \cite{Chenoilprice}. It achieves this by evaluating features based on the values of neighboring samples. As an extension of the original Relief algorithm, ReliefF is specifically engineered to address its predecessor's shortcomings, offering robust performance when dealing with multi-class problems, noisy data, and missing values \cite{tian2025new}. Its enhanced robustness stems from its practice of averaging information from a set of k nearest neighbors, rather than relying on just a single neighbor.

\vspace{0.25\baselineskip}

The core of the ReliefF algorithm is its weight-updating formula. The weight for a given feature ( A ), denoted as $\text{W}(A)$, is iteratively adjusted based on randomly sampled instances:
\begin{equation}
\text{W}(A) = \text{W}(A) - \sum_{j=1}^{k} \frac{\text{diff}(A, R, H_j)}{mk} + \sum_{C \neq \text{class}(R)}  \left[ \frac{\text{P}(C)}{1-\text{P}(\text{class}(R))} \sum_{j=1}^{k} \frac{\text{diff}(A, R, M_{j}^{(C)})}{mk} \right],
\end{equation}

In this equation, $R$ is a randomly selected instance, $H_j$ is its j-th nearest neighbor from the same class (a "hit"), and $M_{j}^{(C)}$ is its j-th nearest neighbor from a different class $C$(a "miss"). The parameters $k$ and $m$ represent the number of nearest neighbors considered and the total number of sampled instances, respectively. The difference function, $\text{diff}(A, R_1, R_2)$, is calculated as follows:

\begin{equation}
\text{diff}(A, R_1, R_2) = 
\begin{cases}
\frac{|R_1[A] - R_2[A]|}{\text{max}(A) - \text{min}(A)} & \text{if } A \text{ is Continuous} \\
0 & \text{if } A \text{ is Discrete and } R_1[A] = R_2[A] \\
1 & \text{if } A \text{ is Discrete and } R_1[A] \neq R_2[A]
\end{cases}.
\end{equation}

Through these mechanisms, ReliefF effectively overcomes the limitations of the original Relief algorithm concerning multi-class data, feature noise, and incomplete data.

\vspace{0.25\baselineskip}

\subsection{Customized EV-STLLM for EV Charging Forecasting}
While the aforementioned data processing modules provide a rich, multifaceted set of capabilities, effectively adapting large language models to leverage this information for spatiotemporal predictions presents its own set of challenges. A major limitation of standard LLM architectures is their inherent lack of awareness of the explicit spatial topology governing interactions within networks such as those connecting electric vehicle charging stations. Furthermore, naive fine-tuning approaches can result in forgetting valuable pre-trained knowledge. To address these challenges, we propose a tailored prediction module, EV-STLLM. The architecture of this module is designed to synergistically fuse the general representation power of LLMs with domain-specific spatiotemporal knowledge. This is primarily achieved through our novel PFGA module, which directly integrates the charging station network structure into the model’s attention mechanism while strategically retaining the pre-trained weights. To ensure that this complex adaptation remains computationally tractable, we also employ a QLoRA augmented training strategy for parameter fine-tuning. 


\subsubsection{Spatio-Temporal Frequency Embedding and Information Fusion}

To enable an LLM to process complex spatio-temporal data, we must first translate the numerical time series and its associated attributes into a unified, high-dimensional feature space. A standard LLM cannot natively interpret the distinct spatial, temporal, and value-based characteristics of EV charging data. Therefore, the primary goal of this module is to create a rich representation for each data point by generating distinct embeddings for the EV charging data's core value (token), its temporal context (time), and its spatial origin (zone), before fusing them into a single comprehensive input for the main forecasting block.

\vspace{0.25\baselineskip}

To begin, we treat the time series at each spatial zone as a sequence of distinct tokens. We then generate the following parallel embeddings:

\vspace{0.25\baselineskip}
~\textbf{Token Embedding ($E_P$)}: To transform the raw numerical EV charging data into a representation that is semantically understandable to the LLM, we first generate token embeddings. This is crucial as it projects the simple scalar values into a high-dimensional space where the model can learn a richer, more abstract understanding of the charging volume or occupancy itself. This is achieved via a pointwise (1x1) convolution:
\begin{equation}
    \mathbf{E}_P = PConv(\mathbf{X}_P; \theta_p),
\end{equation}
where $PConv$ designates a $1 \times 1$ convolution with parameters $\theta_p$, and $D$ is the embedding dimension.

\vspace{0.25\baselineskip}

\textbf{Temporal Embedding ($E_T$)}: Concurrently, to encode the crucial cyclical patterns inherent in EV charging behavior, we create temporal embeddings. We assign absolute positional encodings at daily ($X_{day}$) and weekly ($X_{week}$) granularities and use learnable linear projections to generate embeddings for hour-of-day and day-of-week. This allows the model to explicitly capture periodic trends:
\begin{equation}
    \mathbf{E}_T^{d} = \mathbf{W}_d(\mathbf{X}_{day}),
\end{equation}
\begin{equation}
    \mathbf{E}_T^{w} = \mathbf{W}_w(\mathbf{X}_{week}),
\end{equation}
\begin{equation}
    \mathbf{E}_T = \mathbf{E}_T^{d} + \mathbf{E}_T^{w},
\end{equation}
with $\mathbf{W}_d \in \mathbb{R}^{T_d \times D}$ and $\mathbf{W}_w \in \mathbb{R}^{T_w \times D}$ as learnable parameter matrices. Summing the two yields the overall temporal embedding $\mathbf{E}_T$.

\vspace{0.25\baselineskip}

~\textbf{Spatial Embedding ($E_S$)}: To account for the unique, location-specific factors influencing charging demand (e.g., a station's proximity to a residential versus a commercial area), we learn an adaptive spatial embedding for each station. This provides the model with a distinct signature for each location, enabling it to model geographically diverse patterns:
\begin{equation}
    \mathbf{E}_S = \sigma(\mathbf{W}_S \cdot \mathbf{X}_P + \mathbf{b}_S),
\end{equation}
where $\sigma$ is an activation function, and $\mathbf{W}_S$, $\mathbf{b}_S$ are trainable parameters.

\vspace{0.25\baselineskip}

Finally, a fusion convolution $FConv$ then concatenates and projects the token, spatial, and temporal embeddings into a unified feature:
\begin{equation}
    \mathbf{H}_F = FConv(\mathbf{E}_P \| \mathbf{E}_S \| \mathbf{E}_T; \theta_f),
\end{equation}
where $\mathbf{H}_F \in \mathbb{R}^{N \times 3D}$, $\|$ denotes concatenation, and $\theta_f$ are the fusion parameters.

\subsubsection{Partially Frozen Graph Attention Module}
The PFGA LLM is composed of $F+U$ layers: the first $F$ layers have their multi-head attention and feed-forward sublayers frozen to retain pre-trained knowledge, while the last $U$ layers are unfrozen and employ graph-based attention to more effectively encode spatio-temporal dependencies. In these $U$ layers, QLoRA (Quantized Low-Rank Adaptation) is applied to reduce trainable parameters. The final output of the PFGA LLMs is denoted as $\mathbf{H}^L \in \mathbb{R}^{N \times 3D}$. The regression convolutional layer then utilizes $\mathbf{H}^L$ to forecast the next-step EV charging data, represented by $\hat{\mathbf{Y}}_S \in \mathbb{R}^{S \times N \times C}$.

While frozen pre-trained transformers (FPTs) have shown promise in diverse non-language tasks \cite{LuFrozen}, they may overlook both short- and long-term dependencies for prediction. To this end, we propose graph-based attention mechanisms within partially frozen LLMs, termed PFGA LLMs. 
Specifically, we adapt a GPT-2 backbone, freezing the initial $F$ layers and augmenting the final $U$ layers with graph-aware attention using adjacency matrices to explicitly encode spatial relationships. The processing in the initial $F$ frozen layers follows:
\begin{equation}
\begin{split}
\bar{H}^i &= \text{MHA}(\text{LN}(H^i)) + H^i, \\
H^{i+1} &= \text{FFN}(\text{LN}(\bar{H}^i)) + \bar{H}^i,
\end{split}
\end{equation}
where $i$ ranges from 1 to $F-1$, and $H^1 = [H_F + PE]$ with $PE$ as the positional encoding. $\bar{H}^i$ denotes the intermediate output after the frozen multi-head attention and first layer normalization, while $H^{i+1}$ is computed after applying the frozen feed-forward network and second normalization. These operations are defined as:
\begin{equation}
\begin{split}
\text{LN}(H^i) &= \gamma \odot \frac{H^i - \mu}{\sigma} + \beta, \\
\text{MHA}(\tilde{H}^i) &= W^O(\text{head}_1 || \cdots || \text{head}_h), \\
\text{head}_i &= \text{Attention}(W^Q_i \tilde{H}^i, W^K_i \tilde{H}^i, W^V_i \tilde{H}^i), \\
\text{Attention}(\tilde{H}^i) &= \text{softmax}\left(\frac{\tilde{H}^i\tilde{H}^{iT}}{\sqrt{d_k}}\right)\tilde{H}^i, \\
\text{FFN}(\hat{H}^i) &= \max(0, W_1\hat{H}_P^i + b_1)W_2 + b_2,
\end{split}
\end{equation}
with $\gamma$ and $\beta$ as normalization parameters, and $\odot$ indicating element-wise multiplication.

\vspace{0.25\baselineskip}

Within the final $U$ layers, the adjacency matrix is employed as an attention mask, thus each node can attend to its temporal predecessors and spatial neighbors in accordance with the network topology. Specifically, the multi-head attention modules are unfrozen and modified to incorporate the graph adjacency matrix $A$:
\begin{equation}
\begin{split}
\bar{\mathbf{H}}^{F+U-1} &= \text{MHA}(\text{LN}(\mathbf{H}^{F+U-1}), A) + \mathbf{H}^{F+U-1}, \\
\mathbf{H}^{F+U} &= \text{FFN}(\text{LN}(\bar{\mathbf{H}}^{F+U-1})) + \bar{\mathbf{H}}^{F+U-1},
\end{split}
\end{equation}
where the adjacency matrix $A$ restricts attention to spatially proximate zones, thereby enabling the model to capture complex spatio-temporal interdependencies within the EV charging data. $\mathbf{H}^{F+U}$ constitutes the output after the last layer.

\subsubsection{QLoRA-augmented Training Strategy}

To further enhance computational efficiency and reduce memory overhead, we incorporate QLoRA \cite{goswami2024parameter}, a more advanced fine-tuning strategy than the standard LoRA. QLoRA enables fine-tuning with significantly less memory by backpropagating gradients through a frozen, 4-bit quantized pre-trained language model into Low-Rank Adapters. This approach makes it feasible to fine-tune very large models on limited hardware without sacrificing performance.
The core innovation of QLoRA can be divided into two parts. First, it introduces a new data type, a 4-bit NormalFloat (NF4), which is information-theoretically optimal for normally distributed weights. The weights of the pre-trained model within the PFGA LLMs are quantized to this NF4 format and remain frozen during training. Second, QLoRA uses Double Quantization, a technique that further reduces the memory footprint by quantizing the quantization constants themselves.

While the full pre-trained weight matrices ($W_{i}^{Q}$, $W_{i}^{K}$, and $W_{i}^{V}$) are quantized and frozen, the adaptation is performed by a small set of learnable LoRA parameters. During the forward pass, the 4-bit weights are dequantized to the computation data type (e.g., BFloat16), and the computation proceeds as with a standard LoRA update. We decompose the updates to the query and value matrices into low-rank approximations. Specifically, QLoRA introduces trainable low-rank matrices $L_{i}^{Q}\in\mathbb{R}^{3D\times r}$ and $M_{i}^{Q}\in\mathbb{R}^{r\times d_{k}}$, with $r$ being the rank of the approximation ($r\ll d_{k}$) as follows:
\begin{equation}
\Delta W_{i}^{Q}=L_{i}^{Q}M_{i}^{Q}, \quad \Delta W_{i}^{V}=L_{i}^{V}M_{i}^{V}.
\end{equation}

The effective weight matrices for the query and value are dynamically composed during the forward pass by adding the dequantized base weights and the adapter weights. Let $dQ(\cdot)$ denote the dequantization function from 4-bit to the higher-precision computation format. The updated weights ${W_{i}^{\prime}}^{Q}$ and ${W_{i}^{\prime}}^{V}$ are represented as:
\begin{equation}
{W_{i}^{\prime}}^{Q}=dQ(W_{i, quant}^{Q})+\Delta W_{i}^{Q}, \quad {W_{i}^{\prime}}^{V}=dQ(W_{i, quant}^{V})+\Delta W_{i}^{V}.
\end{equation}

With QLoRA applied, the attention mechanism is updated with these dynamically composed weights, allowing for efficient yet effective fine-tuning:
\begin{align}
MHA(\tilde{H}^{i}) &= W^{O}(head_{1}^{qlora}||\cdot\cdot\cdot||head_{h}^{qlora}), \nonumber \\
head_{i}^{qlora} &= \text{Attention}(W_{i}^{Q^{\prime}}\tilde{H}^{i}, dQ(W_{i, quant}^{K})\tilde{H}^{i},W_{i}^{V^{\prime}}\tilde{H}^{i}).
\end{align}

This strategy dramatically lowers the number of trainable parameters and memory usage, thereby improving scalability and enabling the model to adapt to the specific graph structure of traffic data with exceptional efficiency.

\subsection{Customized Time-Frequency Fusion Loss Function}
Traditional forecasting models typically rely on loss functions that operate exclusively in the time domain, such as Mean Absolute Error. While effective for minimizing point-wise errors, this approach has a notable limitation in direct multi-step forecasting: it may overlook the intrinsic autocorrelation within the label sequence. To this end, we propose a Customized Time-Frequency Fusion Loss Function designed to address this issue. Specifically, our approach first computes the standard loss in the time domain to ensure predictive accuracy. It then performs a secondary alignment in the frequency domain by minimizing the difference between the Fourier transforms of the predicted and true sequences.

\subsubsection{Time Domain Loss Component}

Accurate forecasting of EV charging loads is vital for maintaining the stability of power grids, optimizing the allocation of charging infrastructure, and ensuring user satisfaction through reliable charger availability. The time-domain element of our loss function aims to minimize the discrepancy between the predicted and actual charging demands over time. Specifically, we employ the Mean Absolute Error (MAE) metric, which calculates the average absolute difference between predictions and ground truth values. Compared to Mean Squared Error (MSE), MAE is less influenced by extreme values, making it more suitable for the inherently volatile nature of EV charging data. The MAE is mathematically expressed as:
\begin{equation}
MAELoss = \frac{1}{m}\sum_{i=1}^{m}|\hat{Y}_{i}-Y_{i}|,
\end{equation}
where $m$ represents the total number of predictions, $\hat{Y}_{i}$ denotes the forecasted charging demand, and $Y_{i}$ is the corresponding observed value.

\subsubsection{Frequency Domain Loss Component}

Long-term sequence prediction strategies generally fall into two categories: iterative forecasting (IF) and direct forecasting (DF). IF generates future values step by step, which can result in the accumulation of errors over time. DF, in contrast, predicts the entire sequence in one shot, avoiding error propagation but potentially overlooking the intrinsic autocorrelation present in the sequence, thereby limiting prediction fidelity.

\vspace{0.25\baselineskip}
To mitigate the limitations associated with direct forecasting, our method leverages the Fast Fourier Transform (FFT) to project temporal sequences into the frequency domain. This transformation decomposes the series into its constituent frequencies, reducing temporal dependencies and highlighting periodic patterns. For a time series $x = [x_0, \ldots, x_{T-1}]$, the discrete Fourier transform is defined as:

\begin{equation}
x_k^{(F)} = \sum\limits_{t = 0}^{T - 1} {{x_t}\exp \left( { - j\left( {2\pi /T} \right)kt} \right), \quad \quad 0 \le k \le T - 1} ,
\end{equation}
where $j$ is the imaginary unit, and $x_k^{(F)}$ represents the frequency component at index $k$. For computational efficiency, we utilize the FFT algorithm, denoted as $\mathcal{F}(x)$. The frequency-domain loss quantifies the absolute difference between the frequency spectra of the predicted and true sequences as:

\begin{equation}
FrequencyLoss = \frac{1}{m} \sum_{i=1}^m \left| \mathcal{F}(\hat{Y}_i) - \mathcal{F}(Y_i) \right|,
\end{equation}
where $\mathcal{F}(\cdot)$ is the FFT operator, $Y_i$ is the ground truth sequence for the $i$-th instance, and $\hat{Y}_i$ is the corresponding model prediction.

By integrating the time-domain and frequency-domain loss terms, the proposed loss function enables the model to simultaneously learn from both temporal dynamics and spectral features, thereby improving the robustness and precision of long-term EV charging demand forecasts. The overall fused loss is formulated as follows:

\begin{equation}
CustomizedLoss = MAELoss + \lambda \cdot FrequencyLoss,
\end{equation}
where $\lambda$ is a tunable hyperparameter that controls the relative weight of the frequency-domain loss term.

\section{Data Selection, Parameter Design, and Evaluation Index}

In this section, we primarily introduce datasets from different periods, sourced from the Shenzhen EV charging datasets \cite{Li2025UrbanEV}, the hyperparameters and internal parameters of the data preprocessing module and the EV-STLLM, and the evaluation metrics utilized for a comprehensive assessment of the model's performance.

\subsection{Data selection}
Our study is based on a public dataset of EV charging records from Shenzhen, China, originally spanning a six-month period from September 2022 to February 2023. The raw data provides information from 1,682 public charging stations (comprising 24,798 piles) at a temporal resolution of five-minute intervals. For our analysis, we utilize a refined version of this data from UrbanEV \cite{Li2025UrbanEV}. This pre-processed dataset aggregates the data into a one-hour temporal interval and focuses on a curated set of 1,362 stations with 17,532 piles. This includes primary charging metrics for each station, namely charging volume, occupancy, and duration, alongside dynamic factors such as time-varying electricity price, service price, and weather conditions. Furthermore, the dataset incorporates crucial spatial attributes like coordinates, adjacency, and distances. Finally, the station's static features include its pile number and station number. Our study focuses on predicting two of the most critical load features: EV charging volume and occupancy. The fundamental characteristics and statistical distributions of these target variables are presented in \textbf{Figure 3} and summarized in \textbf{Table 3}. For our experiments, we randomly selected two distinct zones from this dataset, which we refer to as Data 1 and Data 2.

\begin{figure}[!h]
    \centering
    \captionsetup{labelfont=bf}
    \includegraphics[width=1.0\linewidth]{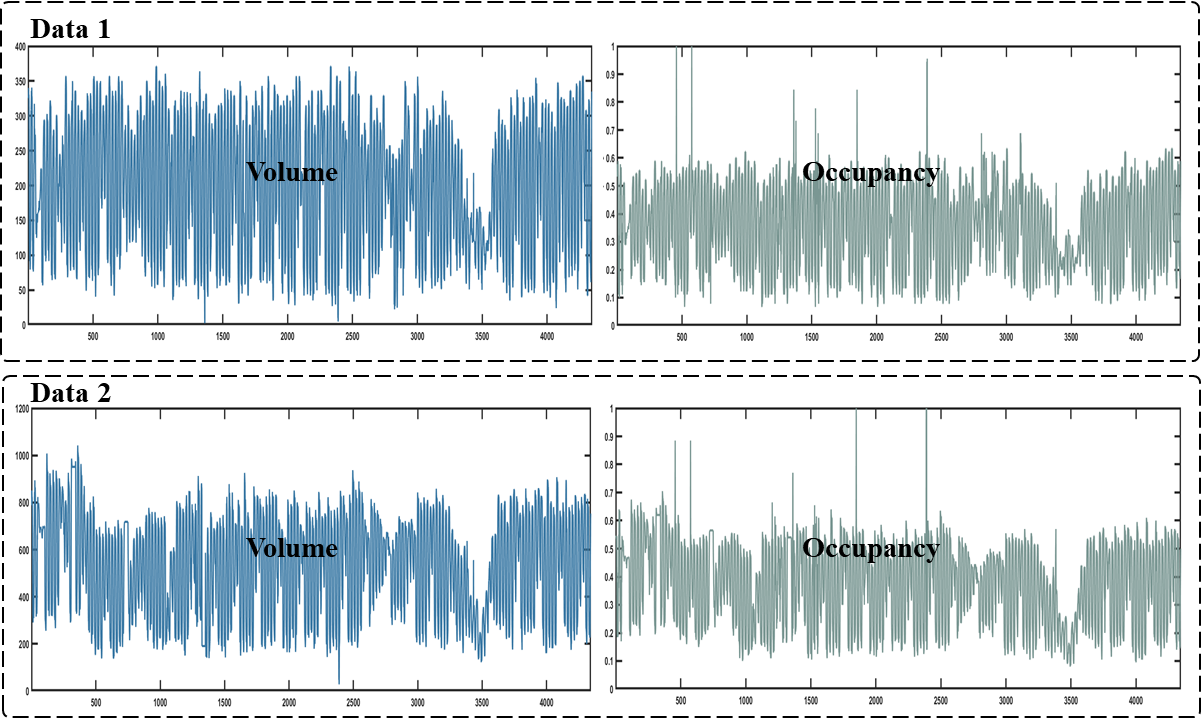}
    \vspace{-0.6\baselineskip}
    \caption{\textbf{The characteristics of volume and occupancy}}
\end{figure}  

\begin{table}[h]
\centering 
\begin{spacing}{1.5}
\caption{\textbf{Basic characteristics of volume and occupancy datasets}}
\small
\resizebox{\textwidth}{!}{
\begin{tabular}{llllllllllll}
\hline
& Tasks     & Data Points & Training Set & Valid Set & Testing Set & Mean   & Max     & Min   & Var      & Std    \\ \hline
\multirow{2}{*}{Data 1} & Volume    & 4345& 80\% & 10\% & 10\% & 198.02 & 371.00  & 2.33  & 9756.64  & 98.78  \\
& Occupancy & 4345 &80\% & 10\% & 10\% & 0.36  & 1.00   & 0.07 & 0.03    & 0.16  \\ 
\hline 
\multirow{2}{*}{Data 2} & Volume & 4345 & 80\% & 10\% & 10\% & 528.99 & 1042.42 & 28.00 & 51023.53 & 225.88 \\
& Occupancy & 4345 & 80\% & 10\% & 10\% & 0.37  & 1.00   & 0.08 & 0.02    & 0.15  \\ \hline
\end{tabular}%
}
\end{spacing}
\end{table}

\subsection{Parameter design} 
The EV-STLLM is divided into two modules: the data processing module and the EV-STLLM forecasting module. In the data processing module, we mainly adopt the VMD-ICEEMDAN technique and the ReliefF algorithm for data preprocessing. In the EV-STLLM forecasting module, we mainly utilize the multi-encoder-single-decoder architecture for accurate forecasting. In this study, we use the Grid Search to optimize the hyper-parameters of VMD-ICEEMDAN and the proposed model. The hyper-parameters of the EV-STLLM are shown in \textbf{Table 4}, and (*) demonstrates the hyper-parameters we adopt in this paper.

\begin{table}[b!]
\begin{spacing}{1.2}
\caption{\textbf{Hyper-parameters of the EV-STLLM}}
\small
\begin{tabularx}{\textwidth}{l l X X}
\toprule 
& \textbf{Model} & \textbf{Parameters} & \textbf{Value} \\
\midrule
\multirow{13}{*}{\textbf{Data Proc.}} & \multirow{5}{*}{\textbf{VMD}} & $\alpha$ & 100 \\
& & $\tau$ & 0 \\
& & $N$ (decomposition layers) & [4, 6, 8*, 10] \\
& & Init & 1 \\
& & Tol & 1e-7 \\
& \multirow{2}{*}{\textbf{ICEEMDAN}} & Ensemble number & 100 \\
& & Noise amplitude & 0.2 \\

& \multirow{3}{*}{\textbf{FIG (FCM)}} & $c$ (Number of granules) & [3, 5*, 7, 9] \\
& & $m$ (fuzziness exponent) & 2 \\
& & Tolerance & 1e-5 \\

& \multirow{2}{*}{\textbf{ReliefF}} & K (number of nearest neighbors) & 70 \\
& & Distance metric & Euclidean \\
\midrule
\multirow{8}{*}{\textbf{Forecasting}} & \multirow{8}{*}{\textbf{EV-STLLM}} & Look-back window & [3, 6, 9, 12*] \\
& & Hidden dimension & 768 \\
& & Time step feature & Based on ReliefF \\
& & Target length & 1, 3, 6, 9 \\
& & Learning rate & [0.00001, 0.0001, 0.001, 0.01*] \\
& & Optimizer & Ranger21 \\
& & Max epoch & [100, 200, 300*, 500] \\
& & Batch size & [32, 64*, 128, 256] \\
\bottomrule
\end{tabularx}
\end{spacing}
\end{table}

\subsection{Evaluation indexes} 
To thoroughly assess the efficacy of our EV-STLLM in EV charging and occupancy forecasting, we employ a suite of evaluation metrics designed to quantify the discrepancy between predicted and true values. Specifically, we utilize MAE, Root Mean Squared Error (RMSE), and Mean Absolute Percentage Error (MAPE), defined as follows:

\begin{equation}
    \text{MAE} = \frac{1}{n} \sum_{i=1}^{n} |\hat{f}_{mi} - f_{mi}|,
\end{equation}

\begin{equation}
    \text{RMSE} = \sqrt{\frac{1}{n} \sum_{i=1}^{n} (\hat{f}_{mi} - f_{mi})^2},
\end{equation}


\begin{equation}
    \text{MAPE} = \frac{1}{n} \sum_{i=1}^{n} \left| \frac{f_{mi} - \hat{f}_{mi}}{f_{mi}} \right| \times 100\%.
\end{equation}



\subsection{Operation environment}
The proposed EV-STLLM operates on a system featuring an Intel i7-9700 CPU @ 3.00 GHz, 16 GB of RAM, and an NVIDIA GeForce RTX 3090. MATLAB 2022a handles data preprocessing. The model was built using PyTorch and developed within the VSCode integrated development environment.

\section{Numerical Verification}

In this section, we conduct four experiments to comprehensively validate the accuracy of our proposed EV-STLLM in EV charging forecasting. Firstly, we validate the superior performance of EV-STLLM with some traditional models and the latest LLM-based forecasting models. Secondly, we conduct an ablation experiment to validate the effectiveness of each component of the EV-STLLM. Moreover, we conduct few-shot studies. Finally, we analyze the robustness of the model through the sensitivity analysis. 

\subsection{Remarkable performance of our proposed frameworks}

In this experimental study, we conducted a comprehensive evaluation of our proposed models against a diverse set of established time series forecasting techniques. The comparison encompassed widely recognized deep learning architectures, namely GCN, LSTM, and PAG \cite{2024PAG}. Furthermore, to underscore the advanced capabilities of our models, we benchmarked their performance against cutting-edge forecasting algorithms based on LLM, including GATGPT, GCNGPT, GPT4TS \cite{Zhou2023}, and ChatEV \cite{qu2024chatev}. To ensure a fair and rigorous assessment, all comparison models were trained using an identical set of features to facilitate optimization, and their respective hyperparameters were meticulously tuned to achieve their optimal performance. 
The visualized forecasting results are shown in \textbf{Figure 4} and \textbf{Figure 5}, and the detailed results are shown in \textbf{Table 5}.

\begin{figure}[!h]
    \centering
    \captionsetup{labelfont=bf}
    \includegraphics[width=0.96\linewidth]{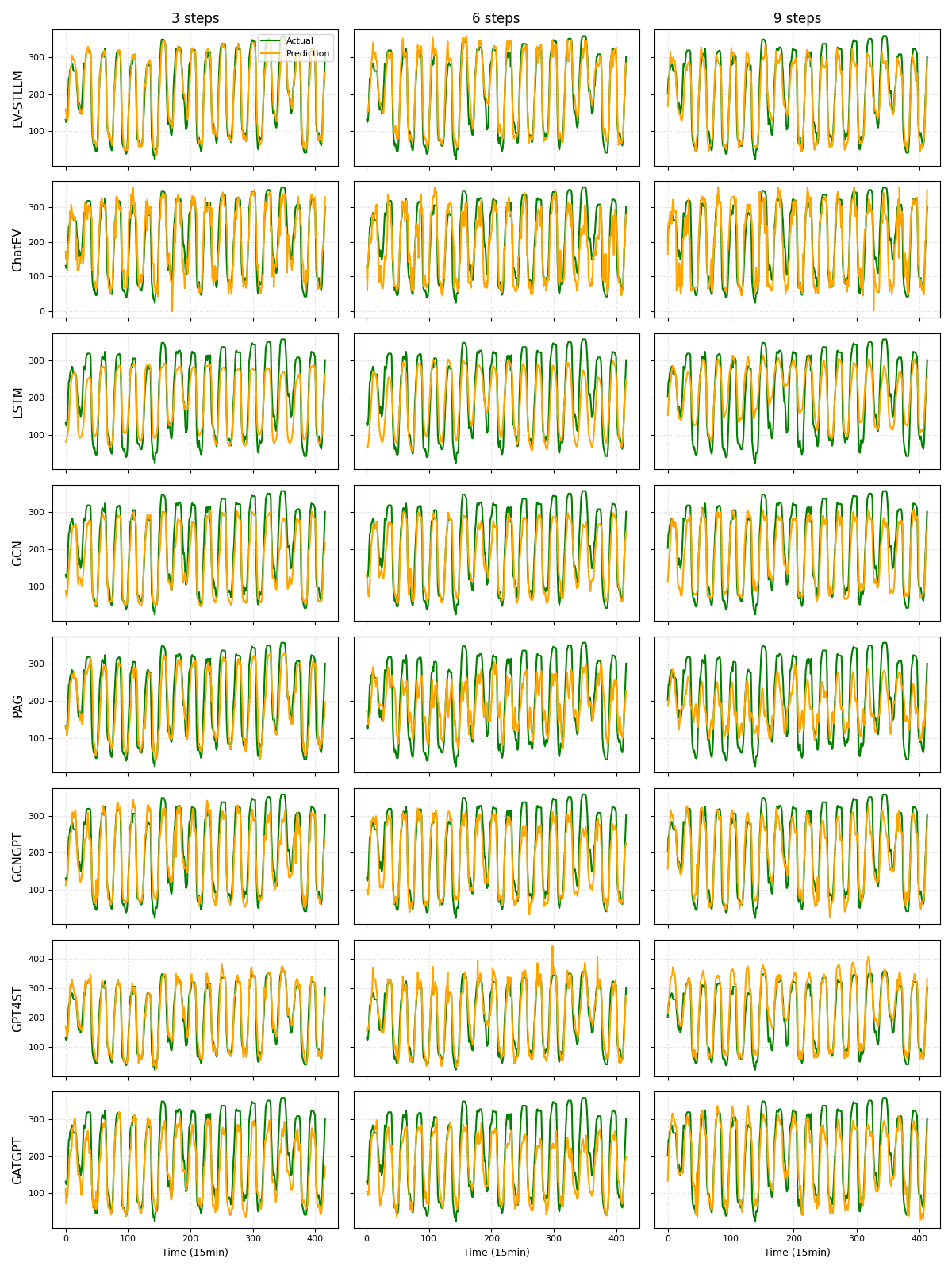}
    \vspace{-0.6\baselineskip}
    \caption{\textbf{Prediction results of different models on volume}}
\end{figure}

\begin{figure}[!h]
    \centering
    \captionsetup{labelfont=bf}
    \includegraphics[width=0.96\linewidth]{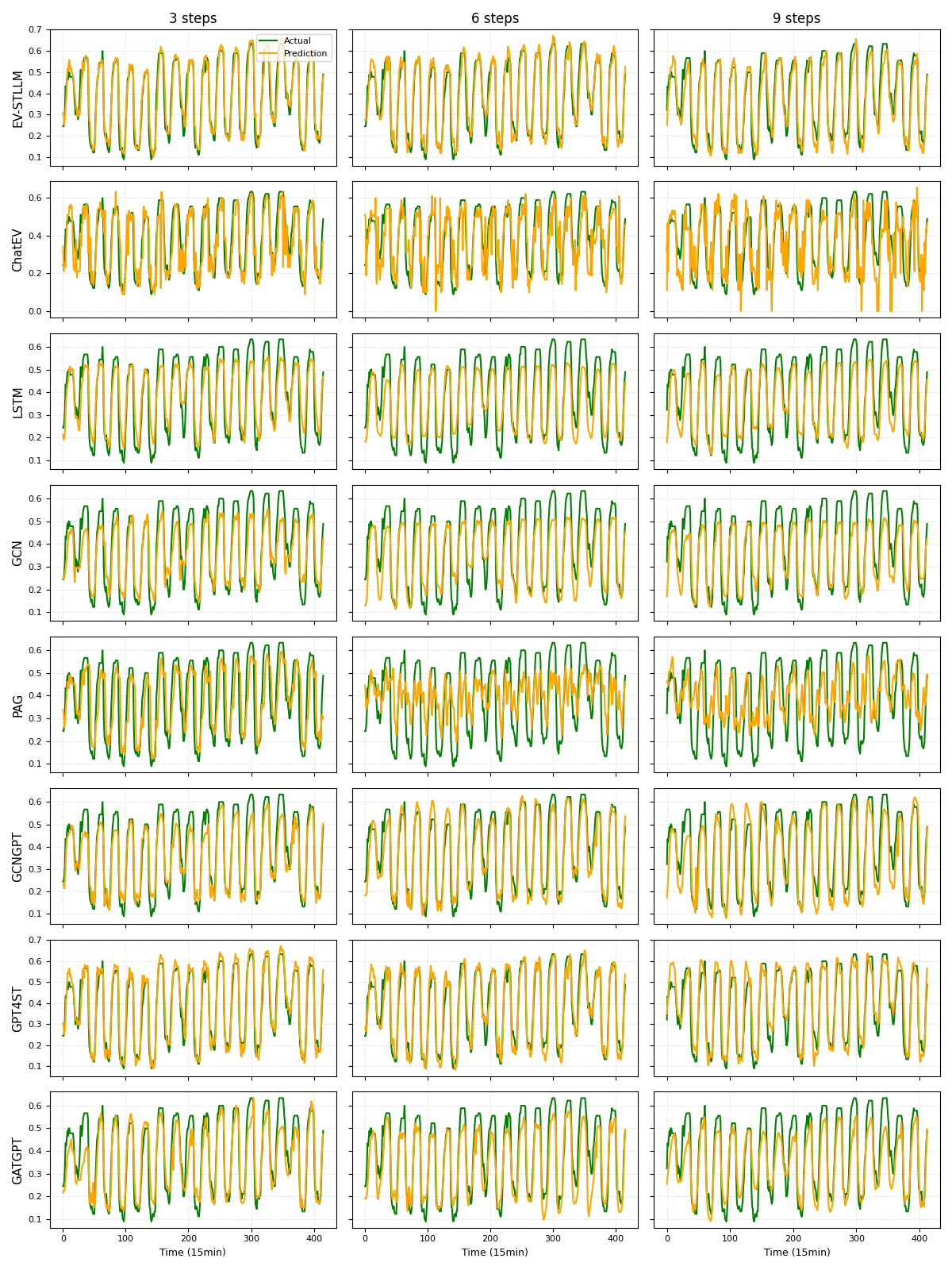}
    \vspace{-0.6\baselineskip}
    \caption{\textbf{Prediction results of different models on occupancy}}
\end{figure}

\vspace{0.25\baselineskip}

On the Data 1 benchmark, our proposed EV-STLLM demonstrates a consistent and significant accuracy advantage over both traditional deep learning architectures and other state-of-the-art forecasting models. For example, in the 3-step volume forecasting task, our model achieved an RMSE of 26.42. This represents a 44.26\% reduction compared to the GCN (RMSE 47.40), highlighting a substantial improvement in predictive reliability. This superior performance also extends to comparisons with other LLM-based methods; our model's MAE of 19.49 represents a 40.81\% improvement over ChatEV's 32.93. Furthermore, when benchmarked against GATGPT in the 6-step volume prediction, EV-STLLM reduces the RMSE by 53.12\%, from 59.26 down to 27.78, cementing its state-of-the-art status.

\vspace{0.25\baselineskip}
The results from Data 2 corroborate these findings, further underscoring the robust capabilities of our framework across different data distributions. In the 3-step task, our model achieves a remarkable 51.56\% reduction in RMSE relative to GCN, with the error declining from 120.95 to 58.59. When evaluated against the GCNGPT architecture, our framework maintains a significant edge, achieving a 46.30\% reduction in MAE from 79.80 to 42.85. Similarly, comparisons with ChatEV show a 46.23\% decrease in RMSE (from 109.53 to 58.89) and a 50\% reduction in MAPE (from 0.18 to 0.09). Even when compared to the highly competitive GPT4TS, our model demonstrates superior performance, reducing the RMSE by 9.75\% (from 65.25 to 58.89) and the MAPE by 25\% (from 0.12 to 0.09). The consistent and significant margins of improvement across both datasets provide unequivocal evidence of our model's robust efficacy and strong generalization capacity in diverse forecasting scenarios.

\vspace{0.25\baselineskip}
In conclusion, the empirical evidence from both Data 1 and Data 2 unequivocally establishes the superior performance of our novel long-sequence forecasting model, EV-STLLM. The framework consistently outperformed a wide spectrum of benchmarks, including established deep learning architectures like GCN, LSTM, and PAG, as well as contemporary LLM-based solutions such as GCNGPT, GATGPT, GPT4TS, and ChatEV. This pronounced and consistent advantage across all key evaluation metrics indicates that EV-STLLM not only delivers a new state-of-the-art in predictive precision but also more effectively mitigates the error accumulation typically observed in extended forecasting horizons.

\vspace{0.25\baselineskip}
\textbf{Remark 1.} Experiment 1 served to validate our proposed EV charging forecasting framework. The results indicate that the framework exhibited excellent performance across all experimental metrics, consistently surpassing other baseline models. Notably, the experimental results across multiple prediction horizons further verified the model's generalization ability and superiority.

\newpage
\newgeometry{landscape, left=1.5cm, right=3cm,top=0.5cm,bottom=0.5cm}
\begin{landscape}
\pdfpagewidth=210mm
\pdfpageheight=297mm
\begin{table}[]
\small
\begin{spacing}{1.2}
\caption{\textbf{Comparison of model performance in two datasets}}
\setlength{\tabcolsep}{0.8mm}
  \begin{tabular}{llllllllllllllllllll}
    \toprule
    \multirow{3}{*}{} & \multirow{3}{*}{} & \multicolumn{6}{l}{3 Steps Prediction} & \multicolumn{6}{l}{6 Steps Prediction} & \multicolumn{6}{l}{9 Steps Prediction} \\
    \cmidrule(lr){3-8} \cmidrule(lr){9-14} \cmidrule(lr){15-20}
    & & \multicolumn{3}{l}{EV Charging} & \multicolumn{3}{l}{Occupancy} & \multicolumn{3}{l}{EV Charging} & \multicolumn{3}{l}{Occupancy} & \multicolumn{3}{l}{EV Charging} & \multicolumn{3}{l}{Occupancy} \\
    \cmidrule(lr){3-5} \cmidrule(lr){6-8} \cmidrule(lr){9-11} \cmidrule(lr){12-14} \cmidrule(lr){15-17} \cmidrule(lr){18-20}
    & & RMSE & MAPE & MAE & RMSE & MAPE & MAE & RMSE & MAPE & MAE & RMSE & MAPE & MAE & RMSE & MAPE & MAE & RMSE & MAPE & MAE \\
    \midrule
    \multirow{8}{*}{Data 1}
    & GCN & 47.40 & 0.20 & 37.96 & 0.07 & 0.20 & 0.06 & 48.42 & 0.24 & 39.17 & 0.08 & 0.23 & 0.07 & 45.54 & 0.24 & 36.95 & 0.08 & 0.21 & 0.07 \\
    & LSTM & 50.83 & 0.30 & 42.43 & 0.07 & 0.21 & 0.06 & 54.79 & 0.31 & 45.33 & 0.08 & 0.24 & 0.07 & 59.07 & 0.45 & 48.32 & 0.08 & 0.22 & 0.06 \\
    & PAG & 56.43 & 0.28 & 41.76 & 0.12 & 0.30 & 0.09 & 67.10 & 0.43 & 56.99 & 0.14 & 0.48 & 0.12 & 75.46 & 0.46 & 63.50 & 0.15 & 0.48 & 0.12 \\
    & GCNGPT & 38.21 & 0.20 & 30.35 & 0.06 & 0.15 & 0.05 & 43.18 & 0.21 & 34.44 & 0.06 & 0.15 & 0.04 & 41.51 & 0.22 & 32.57 & 0.08 & 0.20 & 0.06 \\
    & GATGPT & 53.77 & 0.24 & 43.84 & 0.07 & 0.17 & 0.05 & 59.26 & 0.27 & 48.07 & 0.08 & 0.20 & 0.06 & 46.75 & 0.27 & 37.41 & 0.08 & 0.21 & 0.07 \\
    & GPT4TS & 29.27 & 0.16 & 21.85 & 0.04 & 0.11 & 0.03 & 33.83 & 0.19 & 26.68 & 0.05 & 0.12 & 0.04 & 37.24 & 0.21 & 28.70 & 0.06 & 0.16 & 0.05 \\
    & ChatEV & 44.95 & 0.25 & 32.93 & 0.09 & 0.21 & 0.06 & 57.62 & 0.29 & 44.59 & 0.11 & 0.31 & 0.08 & 54.66 & 0.28 & 39.89 & 0.12 & 0.33 & 0.09 \\
    & \textbf{EV-STLLM} & \textbf{26.42} & \textbf{0.14} & \textbf{19.49} & \textbf{0.04} & \textbf{0.10} & \textbf{0.03} & \textbf{27.78} & \textbf{0.17} & \textbf{21.96} & \textbf{0.04} & \textbf{0.10} & \textbf{0.03} & \textbf{34.23} & \textbf{0.17} & \textbf{27.56} & \textbf{0.05} & \textbf{0.14} & \textbf{0.04} \\
    \midrule
    \multirow{8}{*}{Data 2}
    & GCN & 120.95 & 0.17 & 96.16 & 0.07 & 0.18 & 0.05 & 117.24 & 0.22 & 96.67 & 0.07 & 0.21 & 0.06 & 120.95 & 0.17 & 96.16 & 0.07 & 0.23 & 0.06 \\
    & LSTM & 131.87 & 0.25 & 113.06 & 0.06 & 0.17 & 0.05 & 129.66 & 0.27 & 110.16 & 0.07 & 0.22 & 0.06 & 131.87 & 0.25 & 113.06 & 0.07 & 0.25 & 0.06 \\
    & PAG & 138.48 & 0.22 & 100.10 & 0.12 & 0.31 & 0.09 & 149.01 & 0.29 & 124.95 & 0.12 & 0.40 & 0.10 & 138.48 & 0.22 & 100.10 & 0.14 & 0.48 & 0.12 \\
    & GCNGPT & 96.22 & 0.19 & 79.80 & 0.06 & 0.16 & 0.05 & 99.93 & 0.20 & 82.50 & 0.06 & 0.18 & 0.05 & 91.34 & 0.17 & 73.44 & 0.06 & 0.15 & 0.05 \\
    & GATGPT & 128.30 & 0.19 & 104.12 & 0.08 & 0.23 & 0.06 & 106.69 & 0.19 & 86.06 & 0.08 & 0.20 & 0.06 & 128.30 & 0.19 & 104.12 & 0.07 & 0.21 & 0.06 \\
    & GPT4TS & 65.25 & 0.12 & 51.11 & 0.04 & 0.11 & 0.03 & 81.73 & 0.14 & 63.21 & 0.04 & 0.10 & 0.03 & 70.56 & 0.13 & 55.77 & 0.05 & 0.14 & 0.04 \\
    & ChatEV & 109.53 & 0.18 & 76.45 & 0.08 & 0.19 & 0.05 & 143.60 & 0.24 & 104.03 & 0.09 & 0.25 & 0.07 & 109.53 & 0.18 & 76.45 & 0.10 & 0.25 & 0.07 \\
    & \textbf{EV-STLLM} & \textbf{58.89} & \textbf{0.09} & \textbf{42.85} & \textbf{0.04} & \textbf{0.11} & \textbf{0.03} & \textbf{60.36} & \textbf{0.09} & \textbf{46.76} & \textbf{0.04} & \textbf{0.10} & \textbf{0.03} & \textbf{67.85} & \textbf{0.11} & \textbf{52.93} & \textbf{0.05} & \textbf{0.13} & \textbf{0.04} \\
    \bottomrule
  \end{tabular}
\end{spacing}
\end{table}
\end{landscape}
\restoregeometry

\subsection{The ablation experiment for the proposed EV-STLLM}

In this section, we conduct extensive ablation experiments to verify the contribution of each key component within the proposed EV-STLLM. To achieve this, we design six variants of our model by selectively removing or altering specific modules and strategies. The variants are as follows: (1) w/o Multi-frequency Extraction, which removes the VMD-ICEEMDAN module and uses only denoised data; (2) w/o Multi-scale Extraction, which removes the FIG module; (3) w/o Customized Loss Function, which is trained using only the standard time-domain MAE loss; (4) Partially Frozen Attention (PFA), which uses a partially frozen LLM but removes the graph-based attention mechanism to isolate the impact of the spatial graph structure; (5) Full Graph-based Attention (FGA), which applies graph-based attention to all layers of the LLM to assess the effectiveness of our partial freezing strategy; and (6) Full Tuning (FT), which fine-tunes all parameters of the LLM without any frozen layers. The detailed results of these comparisons are presented in \textbf{Table 6}. 

\begin{table}[htbp]
\small
    \centering
    \caption{Ablation experiment performance results at 3 steps}
    \renewcommand\arraystretch{1.2}
    \begin{tabularx}{\textwidth}{ll XXXXXX}
        \toprule
        \multirow{2}{*}{} & \multirow{2}{*}{} 
        & \multicolumn{3}{l}{EV Charging (Volume)} 
        & \multicolumn{3}{l}{Occupancy} \\
        \cmidrule(lr){3-5} \cmidrule(lr){6-8}
        & & RMSE & MAPE & MAE & RMSE & MAPE & MAE \\
        \midrule

        \multirow{7}{*}{Data 1}               
            & w/o Multi-frequency Extraction & 28.1732 & 0.1741 & 22.5813 & 0.0382 & 0.1009 & 0.0298 \\
            & w/o Multi-scale Extraction& 28.3037 & 0.1707 & 22.4420 & 0.0383 & 0.1004 & 0.0300 \\
            & w/o Full Tuning & 28.3168 & 0.1709 & 22.3916 & 0.0383 & 0.1001 & 0.0297 \\
            & w/o Full Graph-based Attention & 28.3334 & 0.1738 & 22.3468 & 0.0387 & 0.1017 & 0.0302 \\
            & w/o Partially Frozen Attention & 28.4477 & 0.1725 & 22.3167 & 0.0387 & 0.1001 & 0.0302 \\
            & w/o Customized Loss Function& 28.6579 & 0.1750 & 22.6631 & 0.0393 & 0.1039 & 0.0308 \\
            & \textbf{EV-STLLM} & \textbf{27.7834} & \textbf{0.1694} & \textbf{21.9644} & \textbf{0.0378} & \textbf{0.0985} & \textbf{0.0295} \\
            \midrule
        \multirow{7}{*}{Data 2} 
            & w/o Multi-frequency Extraction & 61.2205 & 0.0981 & 47.9574 & 0.0375 & 0.1017 & 0.0282 \\
            & w/o Multi-scale Extraction& 61.4753 & 0.0993 & 47.8945 & 0.0375 & 0.1020 & 0.0282 \\
            & w/o Full Tuning & 61.6983 & 0.1012 & 48.8033 & 0.0376 & 0.1024 & 0.0283 \\
            & w/o Full Graph-based Attention & 62.2325 & 0.1008 & 48.5521 & 0.0378 & 0.1016 & 0.0284 \\
            & w/o Partially Frozen Attention & 62.5775 & 0.1010 & 49.0573 & 0.0379 & 0.1044 & 0.0288 \\
            & w/o Customized Loss Function& 63.5547 & 0.1024 & 49.9255 & 0.0380 & 0.1020 & 0.0284 \\
            & \textbf{EV-STLLM} & \textbf{60.3626} & \textbf{0.0976} & \textbf{46.7596} & \textbf{0.0373} & \textbf{0.1040} & \textbf{0.0286} \\
            \midrule
    \end{tabularx}
\end{table}

\vspace{0.25\baselineskip}
The results indicate that each component plays a crucial role in the model's overall performance, as the removal of any single element leads to a degradation in accuracy across both datasets. The most significant performance drop is observed in the w/o Customized Loss Function variant. On Data 2, removing our combined time-and-frequency domain loss and relying only on a standard loss function caused the RMSE for volume prediction to increase from 60.3626 to 63.5547, and the MAE from 46.76 to 49.93. This highlights the substantial benefit of optimizing the model in both the temporal and spectral domains simultaneously. The importance of our data preprocessing strategy is also evident. Removing the Multi-frequency Extraction module (w/o Multi-frequency Extraction) resulted in a notable increase in error, with the MAE on Data 1 rising from 21.9644 to 22.5813. This confirms the value of decomposing the signal with VMD-ICEEMDAN to capture dynamics at different frequencies. Similarly, the absence of the Multi-scale Extraction module led to performance degradation, validating the effectiveness of using FIG to create features at different temporal granularities. Furthermore, the experiments validate our specific architectural choices for the LLM fine-tuning. The w/o Partially Frozen Attention variant, which is analogous to a model without the graph-based attention mechanism, performed worse than the full model (e.g., RMSE on Data 1 increased from 27.7834 to 28.3334), directly proving the efficacy of integrating the spatial graph structure. Moreover, the variants representing Full Tuning and Full Graph-based Attention also yielded higher errors than our proposed partial approach. This demonstrates that our strategy of selectively freezing initial layers while applying graph attention only to the final, unfrozen layers strikes an optimal balance, preserving essential pre-trained knowledge while effectively adapting the model to the specific spatio-temporal dependencies of the EV charging domain.

\vspace{0.25\baselineskip}

\textbf{Remark 2.} Experimental results validate the contributions of EV-STLLM’s components, highlighting the data preprocessing architecture and the customized time-frequency loss function as particularly instrumental in achieving the model’s enhanced forecasting performance.

\subsection{Evaluation of few-shot learning capability}

In the real world, acquiring large datasets for training models is often impractical and cost-prohibitive. Therefore, a model's ability to achieve high accuracy with only a minimal amount of training data is critical. Large Language Models, benefiting from the vast knowledge encoded during their extensive pre-training, are particularly noted for their strong performance in these situations. To assess our proposed model, we designed a few-shot experiment. In this setup, we evaluate the performance of our proposed model and all baselines after training them on the same training data. The objective is to quantify the robustness and rapid adaptability when faced with limited data availability. Our few-shot learning results are summarized in \textbf{Table 7} and \textbf{Figure 6}. 
\begin{figure}[!h]
    \centering
    \captionsetup{labelfont=bf}
    \includegraphics[width=0.9\linewidth]{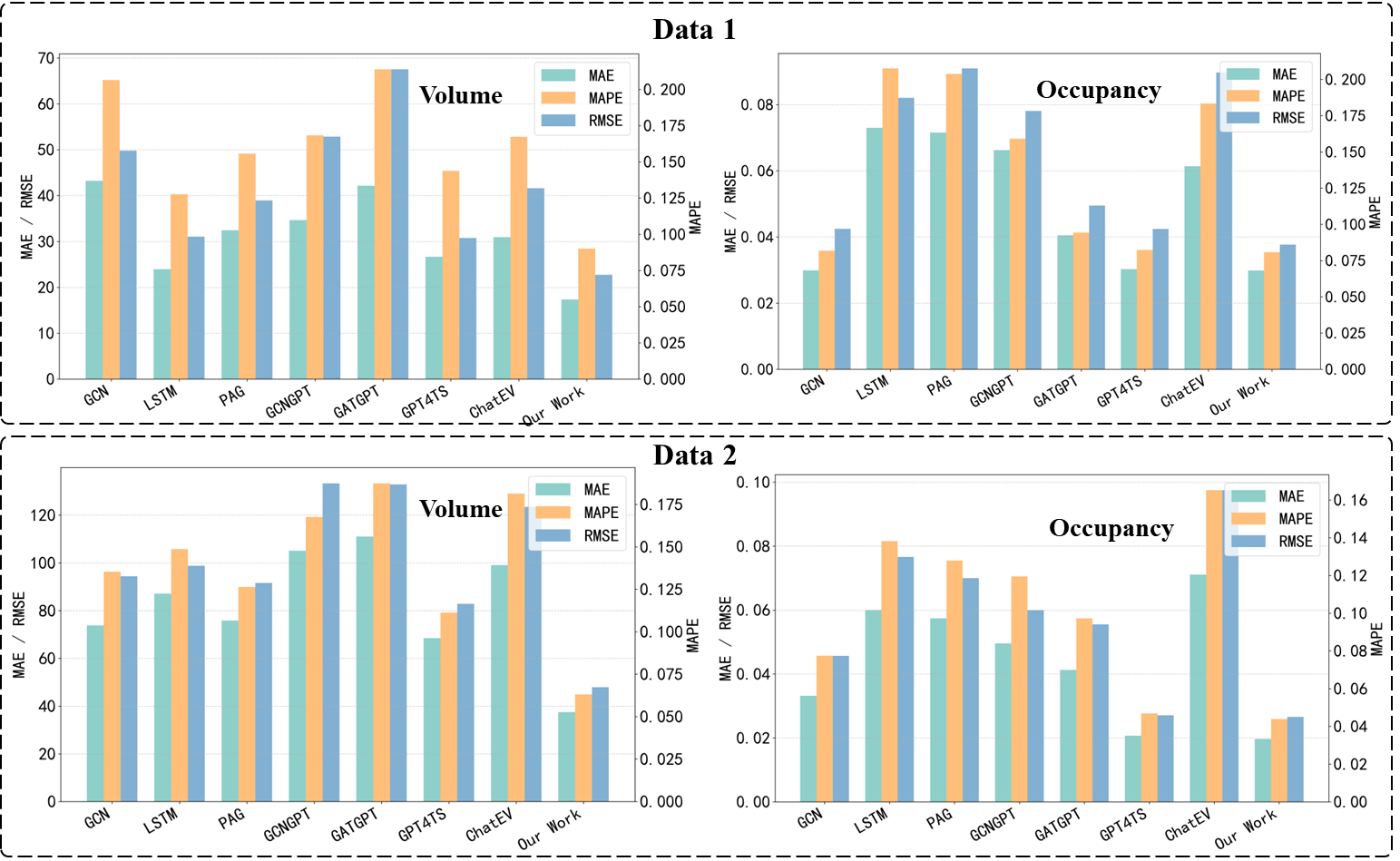}
    \vspace{-0.2\baselineskip}
    \caption{\textbf{The few-shot results of EV-STLLM}}
\end{figure}

\begin{table}[H]
\small
 \caption{Few-shot experiment prediction results at 1 step}
 \renewcommand\arraystretch{1.2}
 \begin{tabularx}{\textwidth}{ll XXXXXX}
  \toprule
  \multirow{2}{*}{} & \multirow{2}{*}{} 
  & \multicolumn{3}{l}{EV Charging (Volume)} 
  & \multicolumn{3}{l}{Occupancy} \\
  \cmidrule(lr){3-5} \cmidrule(lr){6-8}
  & & RMSE & MAPE & MAE & RMSE & MAPE & MAE \\
  \midrule

  \multirow{8}{*}{Data 1} & GCN & 49.7831 & 0.2065 & 43.2456 & 0.0425 & 0.0818 & 0.0299 \\
  & LSTM & 31.0725 & 0.1276 & 23.9484 & 0.0821 & 0.2076 & 0.0730 \\
  & PAG & 38.9430 & 0.1557 & 32.4544 & 0.0910 & 0.2038 & 0.0716 \\
  & GCNGPT & 52.8642 & 0.1684 & 34.6633 & 0.0781 & 0.1591 & 0.0663 \\
  & GATGPT & 67.5307 & 0.2139 & 42.1733 & 0.0495 & 0.0943 & 0.0405 \\
  & GPT4TS & 30.7886 & 0.1438 & 26.6694 & 0.0424 & 0.0824 & 0.0303 \\
  & ChatEV & 41.6304 & 0.1673 & 30.9385 & 0.0897 & 0.1835 & 0.0614 \\
  & \textbf{EV-STLLM} & \textbf{22.7627} & \textbf{0.0901} & \textbf{17.3458} & \textbf{0.0377} & \textbf{0.0808} & \textbf{0.0298} \\
  \midrule
  \multirow{8}{*}{Data 2} & GCN & 94.4019 & 0.1355 & 73.8139 & 0.0457 & 0.0775 & 0.0332 \\
  & LSTM & 98.8168 & 0.1487 & 87.0957 & 0.0766 & 0.1382 & 0.0600 \\
  & PAG & 91.5729 & 0.1264 & 75.8236 & 0.0700 & 0.1279 & 0.0574 \\
  & GCNGPT & 133.3145 & 0.1676 & 105.1271 & 0.0600 & 0.1195 & 0.0496 \\
  & GATGPT & 132.9195 & 0.1874 & 111.0880 & 0.0555 & 0.0973 & 0.0413 \\
  & GPT4TS & 82.8669 & 0.1113 & 68.5098 & 0.0271 & 0.0469 & 0.0207 \\
  & ChatEV & 123.4451 & 0.1814 & 99.1032 & 0.0975 & 0.1652 & 0.0711 \\
  & \textbf{EV-STLLM} & \textbf{47.9232} & \textbf{0.0631} & \textbf{37.5144} & \textbf{0.0266} & \textbf{0.0439} & \textbf{0.0197} \\

  \bottomrule
 \end{tabularx}
\end{table}

\vspace{0.25\baselineskip}

On Data 2, the advantage is even more pronounced. Our model achieves an EV Charging MAE of 37.5144, which constitutes a 45.24\% improvement over the strongest LLM baseline, GPT4TS (MAE of 68.5098), and a remarkable 62.15\% improvement over ChatEV (MAE of 99.1032). This substantial margin underscores the effectiveness of our specialized architecture. While general-purpose LLM forecasters like GPT4TS leverage pre-trained knowledge, our model's integration of a sophisticated data processing module and a graph-aware attention mechanism allows it to adapt this knowledge to the specific domain much more efficiently. Compared to traditional deep learning models, which lack pre-trained knowledge, the performance gap is vast. For instance, on Data 1, our model's MAE of 17.3458 for EV Charging is 59.89\% lower than GCN's 43.2456. Similarly, on Data 2, our model's RMSE of 47.9232 is 51.50\% lower than LSTM's 98.8168. This indicates that traditional models struggle significantly to learn meaningful patterns from scarce data, whereas our LLM-based framework can rapidly generalize.

\vspace{0.25\baselineskip}
To summarize, the experiment results unequivocally demonstrate that the time-frequency domain feature learning strategy significantly boosts the model's capacity to discern local features within long-sequence data. Furthermore, the tailored loss function facilitates a seamless integration of time-domain and frequency-domain insights, thereby refining the model's multi-frequency information representation capabilities.

\vspace{0.25\baselineskip}
\textbf{Remark 3.} These results highlight a key insight for applying LLMs to specialized domains. LLMs’ inherent pre-trained knowledge provides a powerful foundation, particularly useful for few-shot learning. However, achieving state-of-the-art performance crucially relies on targeted, domain-specific enhancements, such as our advanced data processing and graph-aware attention.

\subsection{The forecasting performance of the EV-STLLM for critical days}
The distinct data patterns observed on critical days pose a significant challenge to achieving high prediction accuracy. Inaccurate forecasts for these periods can result in substantial resource mismatches for grid operators and charging service providers alike. Existing forecasting models mainly focus on the prediction of regular days \cite{CriticalDayMatters}. To this end, we propose the inclusion of a binary indicator variable (0 or 1) to enhance the model's predictive accuracy on critical days.
In this study, we select five days during the Chinese Spring Festival holiday (2023.1.22-2023.1.26) as representative critical days. The five consecutive workdays immediately following this holiday were designated as working days. Our training set spans from September 1, 2022, to January 16, 2023, with the validation set covering January 17-21, 2023. \textbf{Figure 7} visually illustrates the notable difference in charging behavior between these two periods.

\begin{figure}[!h]
    \centering
    \captionsetup{labelfont=bf}
    \includegraphics[width=1.0\linewidth]{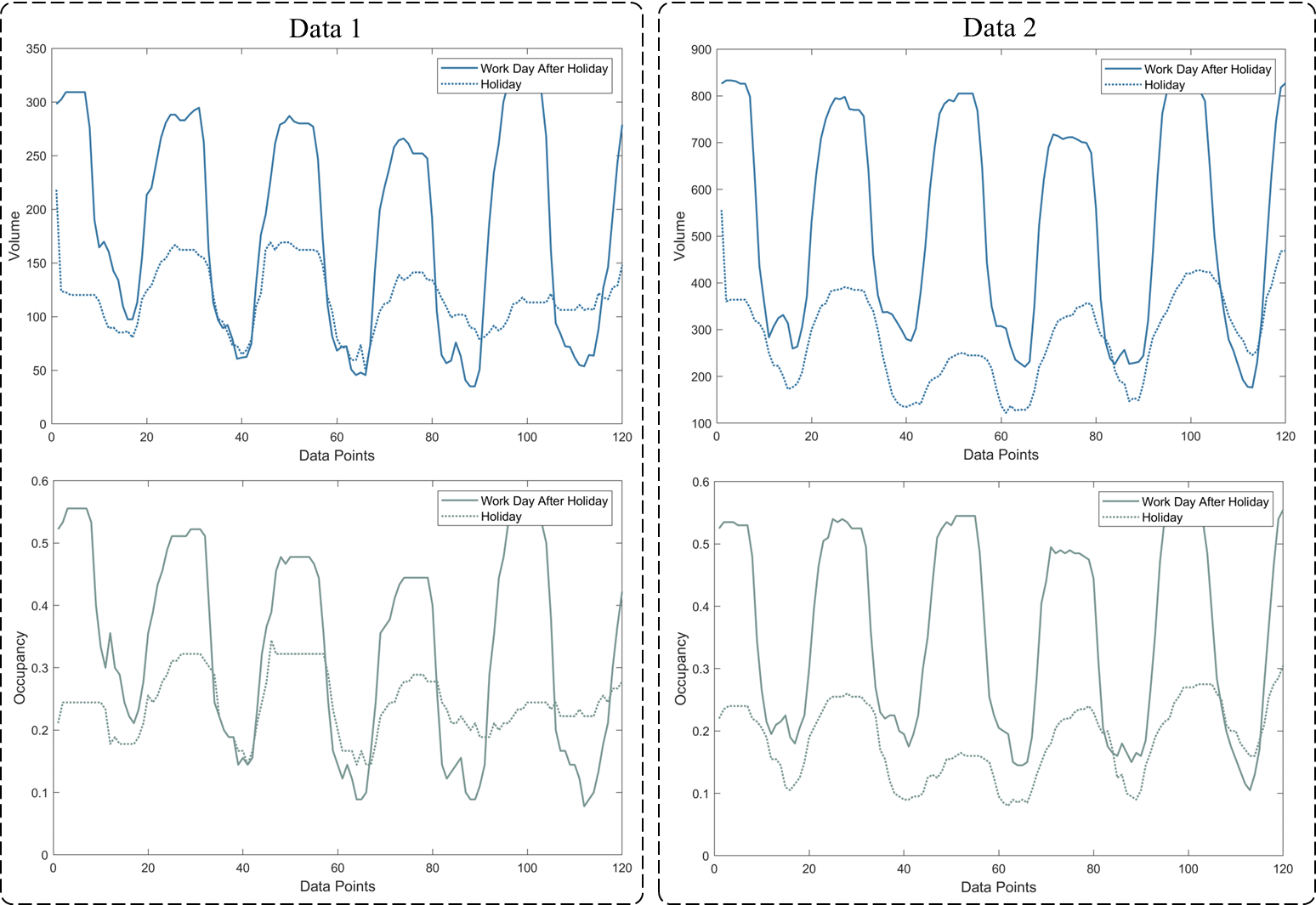}
    \vspace{-0.6\baselineskip}
    \caption{\textbf{The critical day of volume and occupancy}}
\end{figure}  

As illustrated in \textbf{Figure 7}, a pronounced discrepancy exists between charging patterns on holidays and the subsequent workdays across both Data 1 and Data 2. The data for post-holiday workdays (solid line) displays a highly regular and cyclical pattern for both charging volume and station occupancy, with prominent peaks that likely correspond to daily commuting and regular activity schedules. In stark contrast, the data from the holiday period (dotted line) shows significantly suppressed and more erratic demand. The distinct daily peaks are largely absent, replaced by a lower, flatter, and less structured demand profile throughout the day.
This stark difference can be attributed to several reasons. First, with most businesses and schools closed, daily commuting—a primary driver of predictable charging demand—ceases almost entirely. Second, as one of the migrant cities in China \cite{TONG2020102082}, many residents travel out of the city or, if they remain, engage in local family activities that do not follow a regular travel schedule. This leads to a significant reduction in the overall use of public charging infrastructure. 

\vspace{0.25\baselineskip}
\begin{table}[H]
\small
\caption{The critical day prediction results at 1 step}
\renewcommand\arraystretch{1.2}
\begin{tabularx}{\textwidth}{ll XXXXXX} 
    \toprule
    \multirow{2}{*}{} & \multirow{2}{*}{} 
    & \multicolumn{3}{l}{EV Charging (Volume)} 
    & \multicolumn{3}{l}{Occupancy} \\
    \cmidrule(lr){3-5} \cmidrule(lr){6-8}
    & & RMSE & MAPE & MAE & RMSE & MAPE & MAE \\
    \midrule

    \multirow{8}{*}{Data 1} 
    & GCN       & 73.7648 & 0.6414 & 68.8062 & 0.1193 & 0.4846 & 0.1101 \\
    & LSTM      & 70.9268 & 0.6571 & 66.7550 & 0.1036 & 0.3943 & 0.0864 \\
    & PAG       & 21.4700 & 0.1207 & 13.6853 & 0.0243 & 0.0742 & 0.0170 \\
    & GCNGPT    & 122.8174 & 1.0385 & 111.8417 & 0.0821 & 0.3104 & 0.0714 \\
    & GATGPT    & 92.3021 & 0.7279 & 79.5813 & 0.0670 & 0.2369 & 0.0542 \\
    & GPT4TS    & 27.6410 & 0.1672 & 19.3018 & 0.0205 & 0.0635 & 0.0145 \\
    & ChatEV    & 62.0898 & 0.4560 & 51.3021 & 0.1044 & 0.3210 & 0.0765 \\
    & \textbf{EV-STLLM} & \textbf{16.2557} & \textbf{0.0777} & \textbf{8.4886} & \textbf{0.0169} & \textbf{0.0535} & \textbf{0.0120} \\

    \midrule

    \multirow{8}{*}{Data 2} 
    & GCN       & 205.7274 & 0.8418 & 195.7456 & 0.2102 & 1.3302 & 0.2022 \\
    & LSTM      & 201.1700 & 0.8390 & 189.9052 & 0.1763 & 1.1282 & 0.1682 \\
    & PAG       & 43.2357 & 0.1035 & 28.4360 & 0.0198 & 0.0891 & 0.0147 \\
    & GCNGPT    & 265.0389 & 1.0297 & 247.4557 & 0.1629 & 1.0080 & 0.1548 \\
    & GATGPT    & 271.6008 & 1.0926 & 251.4485 & 0.1853 & 1.1860 & 0.1805 \\
    & GPT4TS    & 58.3105 & 0.1436 & 40.1546 & 0.0171 & 0.0806 & 0.0126 \\
    & ChatEV    & 131.5461 & 0.3681 & 95.3069 & 0.1152 & 0.5261 & 0.0794 \\
    & \textbf{EV-STLLM} & \textbf{37.1491} & \textbf{0.0833} & \textbf{20.8059} & \textbf{0.0161} & \textbf{0.0784} & \textbf{0.0118} \\

    \bottomrule
\end{tabularx}
\end{table}

The experimental results for the 1-step critical day prediction confirm the exceptional robustness of our EV-STLLM under these challenging conditions. In Data 1, our model achieves an MAE of 8.4886 for volume prediction, a remarkable 37.97\% improvement over the next-best model, PAG (MAE of 13.6853), and a 56.02\% improvement over GPT4TS (MAE of 19.3018). In Data 2, our model achieves an MAE of 20.8059 for volume prediction, a 26.83\% improvement over PAG(MAE of 28.4360), and a 48.19\% improvement over GPT4TS (MAE of 40.1546).

\vspace{0.25\baselineskip}

As demonstrated in \textbf{Table 8}, the EV-STLLM exhibits considerable robustness with minimal sensitivity to variations in critical days, indicating its strong potential for practical engineering applications compared to other models. 

\vspace{0.25\baselineskip}

\textbf{Remark 4.} This experiment assesses the model's generalization capability when faced with critical days. The findings indicate that our proposed model is remarkably stable against drastic pattern shifts, thereby demonstrating its robust performance and reliability for real-world deployment.

\section{Conclusion}

We proposed the EV-STLLM, a novel and robust forecasting framework built upon a spatio-temporal large language model fused with multi-frequency and multi-scale information. Our primary contribution is the development of a comprehensive two-stage framework. The first stage features an advanced data preprocessing module that employs a VMD-ICEEMDAN technique for multi-frequency decomposition and FIG for multi-scale feature extraction. This ensures a deep and nuanced representation of the raw time series data, capturing everything from short-term fluctuations to long-term trends. The second stage introduces a customized spatio-temporal LLM, featuring our innovative PFGA module. By integrating the charging station network topology as an attention mask and selectively fine-tuning layers with QLoRA, the PFGA module effectively adapts the powerful general knowledge of pre-trained LLMs to the specific EV domain, while ensuring computational efficiency. Furthermore, we designed a customized time-frequency fusion loss function that optimizes the model in both domains simultaneously, significantly enhancing predictive accuracy. Our empirical studies show that EV-STLLM consistently outperforms state-of-the-art traffic prediction models and other LLM-based approaches, also proving its robustness in few-shot, zero-shot, and critical day forecasting scenarios. This indicates that our model not only assists with distribution grid energy management but also provides valuable informational support for EV aggregators participating in electricity market transactions.

\vspace{0.25\baselineskip} 

While the proposed framework demonstrates strong performance, future work will focus on extending its capabilities.
First, by utilizing knowledge distillation learning strategies, we can develop lighter-weight models that will improve the localized deployment and predictive power of EV charging stations. Second, integrating prediction results with scheduling strategies to improve the ability of charging stations to participate in electricity market transactions.

\bibliographystyle{elsarticle-num} 
\bibliography{cas-refs}

\begin{thebibliography}{10}
\expandafter\ifx\csname url\endcsname\relax
  \def\url#1{\texttt{#1}}\fi
\expandafter\ifx\csname urlprefix\endcsname\relax\def\urlprefix{URL }\fi
\expandafter\ifx\csname href\endcsname\relax
  \def\href#1#2{#2} \def\path#1{#1}\fi

\bibitem{GUO2015420}
C.~Guo, C.~C. Chan, Analysis method and utilization mechanism of the overall value of ev charging, Energy Conversion and Management 89 (2015) 420--426.
\newblock \href {https://doi.org/https://doi.org/10.1016/j.enconman.2014.10.016} {\path{doi:https://doi.org/10.1016/j.enconman.2014.10.016}}.

\bibitem{fan2025incentive}
H.~Fan, Z.~Li, Y.~Duan, B.~Wang, Incentive policy formulation for china’s electric vehicle market: Navigating pathways to sustainable mobility with a green premium analytical model, Energy Policy 202 (2025) 114610.
\newblock \href {https://doi.org/10.1016/j.enpol.2025.114610} {\path{doi:10.1016/j.enpol.2025.114610}}.

\bibitem{LI2025125933}
H.~Li, Z.~Liu, M.~{Bin Kaleem}, L.~Duan, S.~Ruan, W.~Liu, Fault detection for lithium-ion batteries of electric vehicles with spatio-temporal autoencoder, Applied Energy 392 (2025) 125933.
\newblock \href {https://doi.org/https://doi.org/10.1016/j.apenergy.2025.125933} {\path{doi:https://doi.org/10.1016/j.apenergy.2025.125933}}.

\bibitem{WU2023116619}
Y.~Wu, Z.~Huang, Y.~Zheng, Y.~Liu, H.~Li, Y.~Che, J.~Peng, R.~Teodorescu, Spatial–temporal data-driven full driving cycle prediction for optimal energy management of battery/supercapacitor electric vehicles, Energy Conversion and Management 277 (2023) 116619.
\newblock \href {https://doi.org/https://doi.org/10.1016/j.enconman.2022.116619} {\path{doi:https://doi.org/10.1016/j.enconman.2022.116619}}.

\bibitem{SHANG2025125460}
Y.~Shang, D.~Li, Y.~Li, S.~Li, Explainable spatiotemporal multi-task learning for electric vehicle charging demand prediction, Applied Energy 384 (2025) 125460.
\newblock \href {https://doi.org/https://doi.org/10.1016/j.apenergy.2025.125460} {\path{doi:https://doi.org/10.1016/j.apenergy.2025.125460}}.

\bibitem{WANG2024130305}
S.~Wang, W.~Zhang, Y.~Sun, A.~Trivedi, C.~Y. Chung, D.~Srinivasan, Wind power forecasting in the presence of data scarcity: A very short-term conditional probabilistic modeling framework, Energy 291 (2024) 130305.
\newblock \href {https://doi.org/https://doi.org/10.1016/j.energy.2024.130305} {\path{doi:https://doi.org/10.1016/j.energy.2024.130305}}.

\bibitem{WANG2025134884}
S.~Wang, Y.~Sun, W.~Zhang, D.~Srinivasan, Optimization of deterministic and probabilistic forecasting for wind power based on ensemble learning, Energy 319 (2025) 134884.
\newblock \href {https://doi.org/https://doi.org/10.1016/j.energy.2025.134884} {\path{doi:https://doi.org/10.1016/j.energy.2025.134884}}.

\bibitem{Jia2023ReviewOptimizationForecasting}
Z.~Jia, J.~Li, X.-P. Zhang, R.~Zhang, Review on optimization of forecasting and coordination strategies for electric vehicle charging, Journal of Modern Power Systems and Clean Energy 11~(2) (2023) 389--400.
\newblock \href {https://doi.org/10.35833/MPCE.2021.000777} {\path{doi:10.35833/MPCE.2021.000777}}.

\bibitem{xing2021modelling}
Q.~Xing, Z.~Chen, Z.~Zhang, R.~Wang, T.~Zhang, Modelling driving and charging behaviours of electric vehicles using a data-driven approach combined with behavioural economics theory, Journal of Cleaner Production 324 (2021) 129243.
\newblock \href {https://doi.org/10.1016/j.jclepro.2021.129243} {\path{doi:10.1016/j.jclepro.2021.129243}}.

\bibitem{Li2022ReviewLoadForecastingMethods}
H.~Li, J.~Zhu, Y.~Zhou, D.~Feng, K.~Zhang, B.~Shen, Review of load forecasting methods for electric vehicle charging station, in: 2022 IEEE/IAS Industrial and Commercial Power System Asia (I\&CPS Asia), 2022, pp. 1833--1837.
\newblock \href {https://doi.org/10.1109/ICPSAsia55496.2022.9949707} {\path{doi:10.1109/ICPSAsia55496.2022.9949707}}.

\bibitem{Liu2025}
W.~Liu, M.~Gai, Pv-mlp: A lightweight patch-based multi-layer perceptron network with time–frequency domain fusion for accurate long-sequence photovoltaic power forecasting, Renewable Energy 251, cited by: 0 (2025).
\newblock \href {https://doi.org/10.1016/j.renene.2025.123277} {\path{doi:10.1016/j.renene.2025.123277}}.

\bibitem{WU201552}
X.~Wu, D.~Freese, A.~Cabrera, W.~A. Kitch, Electric vehicles’ energy consumption measurement and estimation, Transportation Research Part D: Transport and Environment 34 (2015) 52--67.
\newblock \href {https://doi.org/https://doi.org/10.1016/j.trd.2014.10.007} {\path{doi:https://doi.org/10.1016/j.trd.2014.10.007}}.

\bibitem{WeicanEnhancing}
W.~Liu, Z.~Tian, Y.~Qiu, Enhancing accuracy in point-interval load forecasting: A new strategy based on data augmentation, customized deep learning, and weighted linear error correction, Expert Systems with Applications 272 (2025) 126686.
\newblock \href {https://doi.org/https://doi.org/10.1016/j.eswa.2025.126686} {\path{doi:https://doi.org/10.1016/j.eswa.2025.126686}}.

\bibitem{8061012}
M.~S. Islam, N.~Mithulananthan, D.~Q. Hung, A day-ahead forecasting model for probabilistic ev charging loads at business premises, IEEE Transactions on Sustainable Energy 9~(2) (2018) 741--753.
\newblock \href {https://doi.org/10.1109/TSTE.2017.2759781} {\path{doi:10.1109/TSTE.2017.2759781}}.

\bibitem{2017Statistical}
Q.~Yan, C.~Qian, B.~Zhang, M.~Kezunovic, Statistical analysis and modeling of plug-in electric vehicle charging demand in distribution systems, in: 2017 19th International Conference on Intelligent System Application to Power Systems (ISAP), 2017, pp. 1--6.
\newblock \href {https://doi.org/10.1109/ISAP.2017.8071365} {\path{doi:10.1109/ISAP.2017.8071365}}.

\bibitem{10690538}
F.~H. Juwono, W.~K. Wong, E.~Purwanto, R.~Reine, H.~Hugeng, Machine learning role in electric vehicles: A review, in: 2024 10th International Conference on Smart Computing and Communication (ICSCC), 2024, pp. 671--675.
\newblock \href {https://doi.org/10.1109/ICSCC62041.2024.10690538} {\path{doi:10.1109/ICSCC62041.2024.10690538}}.

\bibitem{SVM}
M.~Mao, Y.~Yue, L.~Chang, Multi-time scale forecast for schedulable capacity of electric vehicle fleets using big data analysis, in: 2016 IEEE 7th International Symposium on Power Electronics for Distributed Generation Systems (PEDG), 2016, pp. 1--7.
\newblock \href {https://doi.org/10.1109/PEDG.2016.7527051} {\path{doi:10.1109/PEDG.2016.7527051}}.

\bibitem{ANN}
N.~K.~H. Su, F.~H. Juwono, W.~K. Wong, I.~M. Chew, Review on machine learning methods for remaining useful lifetime prediction of lithium-ion batteries, in: 2022 International Conference on Green Energy, Computing and Sustainable Technology (GECOST), 2022, pp. 286--292.
\newblock \href {https://doi.org/10.1109/GECOST55694.2022.10010569} {\path{doi:10.1109/GECOST55694.2022.10010569}}.

\bibitem{LSTM}
Z.~Xu, K.~Cao, Y.~Liu, C.~Wang, Short-term load prediction of ev charging station based on lstm recursion, in: 2024 IEEE 2nd International Conference on Power Science and Technology (ICPST), 2024, pp. 2068--2073.
\newblock \href {https://doi.org/10.1109/ICPST61417.2024.10602175} {\path{doi:10.1109/ICPST61417.2024.10602175}}.

\bibitem{SVM2}
Q.~Sun, J.~Liu, X.~Rong, M.~Zhang, X.~Song, Z.~Bie, Z.~Ni, Charging load forecasting of electric vehicle charging station based on support vector regression, in: 2016 IEEE PES Asia-Pacific Power and Energy Engineering Conference (APPEEC), 2016, pp. 1777--1781.
\newblock \href {https://doi.org/10.1109/APPEEC.2016.7779794} {\path{doi:10.1109/APPEEC.2016.7779794}}.

\bibitem{2024HybridCNN}
A.~Romia, Q.~Huang, A hybrid cnn-lstm-attention deep learning model for forecasting time-series electric vehicles fast charging loads at public stations, in: 2024 IEEE Industry Applications Society Annual Meeting (IAS), 2024, pp. 1--7.
\newblock \href {https://doi.org/10.1109/IAS55788.2024.11023795} {\path{doi:10.1109/IAS55788.2024.11023795}}.

\bibitem{2025HybridARIMA}
M.~I. El-Afifi, A.~A. Eladl, B.~E. Sedhom, M.~A. Hassan, Enhancing ev charging station integration: A hybrid arima-lstm forecasting and optimization framework, IEEE Transactions on Industry Applications 61~(3) (2025) 4924--4935.
\newblock \href {https://doi.org/10.1109/TIA.2025.3540788} {\path{doi:10.1109/TIA.2025.3540788}}.

\bibitem{TIAN2025125174}
J.~Tian, H.~Liu, W.~Gan, Y.~Zhou, N.~Wang, S.~Ma, Short-term electric vehicle charging load forecasting based on tcn-lstm network with comprehensive similar day identification, Applied Energy 381 (2025) 125174.
\newblock \href {https://doi.org/https://doi.org/10.1016/j.apenergy.2024.125174} {\path{doi:https://doi.org/10.1016/j.apenergy.2024.125174}}.

\bibitem{10667370}
X.~Ge, X.~Zhang, D.~Xu, A novel seqgan-lstm load forecasting framework for electric vehicle charging stations with missing data, in: 2024 IEEE 15th International Symposium on Power Electronics for Distributed Generation Systems (PEDG), 2024, pp. 1--6.
\newblock \href {https://doi.org/10.1109/PEDG61800.2024.10667370} {\path{doi:10.1109/PEDG61800.2024.10667370}}.

\bibitem{YIN2025134559}
W.~Yin, J.~Ji, Prediction of ev charging load based on federated learning, Energy 316 (2025) 134559.
\newblock \href {https://doi.org/https://doi.org/10.1016/j.energy.2025.134559} {\path{doi:https://doi.org/10.1016/j.energy.2025.134559}}.

\bibitem{he2016deep}
K.~He, X.~Zhang, S.~Ren, J.~Sun, Deep residual learning for image recognition, in: Proceedings of the IEEE conference on computer vision and pattern recognition, 2016, pp. 770--778.

\bibitem{2024PAG}
H.~Qu, H.~Kuang, Q.~Wang, J.~Li, L.~You, A physics-informed and attention-based graph learning approach for regional electric vehicle charging demand prediction, IEEE Transactions on Intelligent Transportation Systems 25~(10) (2024) 14284--14297.
\newblock \href {https://doi.org/10.1109/TITS.2024.3401850} {\path{doi:10.1109/TITS.2024.3401850}}.

\bibitem{Li2025UrbanEV}
H.~Li, H.~Qu, X.~Tan, L.~You, R.~Zhu, W.~Fan, Urbanev: An open benchmark dataset for urban electric vehicle charging demand prediction, Scientific Data 12 (03 2025).
\newblock \href {https://doi.org/10.1038/s41597-025-04874-4} {\path{doi:10.1038/s41597-025-04874-4}}.

\bibitem{liuSTLLM}
C.~Liu, S.~Yang, Q.~Xu, Z.~Li, C.~Long, Z.~Li, R.~Zhao, Spatial-temporal large language model for traffic prediction, in: 2024 25th IEEE International Conference on Mobile Data Management (MDM), IEEE, IEEE, Brussels, Belgium, 2024, pp. 31--40.
\newblock \href {https://doi.org/10.1109/MDM61037.2024.00025} {\path{doi:10.1109/MDM61037.2024.00025}}.

\bibitem{wang2023predicting}
S.~Wang, A.~Chen, P.~Wang, C.~Zhuge, Predicting electric vehicle charging demand using a heterogeneous spatio-temporal graph convolutional network, Transportation Research Part C: Emerging Technologies 153 (2023) 104205.
\newblock \href {https://doi.org/https://doi.org/10.1016/j.trc.2023.104205} {\path{doi:https://doi.org/10.1016/j.trc.2023.104205}}.

\bibitem{qu2024chatev}
H.~Qu, H.~Li, L.~You, R.~Zhu, J.~Yan, P.~Santi, C.~Ratti, C.~Yuen, Chatev: Predicting electric vehicle charging demand as natural language processing, Transportation Research Part D: Transport and Environment 136 (2024) 104470.
\newblock \href {https://doi.org/10.1016/j.trd.2024.104470} {\path{doi:10.1016/j.trd.2024.104470}}.

\bibitem{ST-LLM+}
C.~Liu, K.~H. Hettige, Q.~Xu, C.~Long, S.~Xiang, G.~Cong, Z.~Li, R.~Zhao, St-llm+: Graph enhanced spatio-temporal large language models for traffic prediction, IEEE Transactions on Knowledge and Data Engineering (2025) 1--14\href {https://doi.org/10.1109/TKDE.2025.3570705} {\path{doi:10.1109/TKDE.2025.3570705}}.

\bibitem{tian2024cnns}
Z.~Tian, W.~Liu, W.~Jiang, C.~Wu, Cnns-transformer based day-ahead probabilistic load forecasting for weekends with limited data availability, Energy 293 (2024) 130666.

\bibitem{sun2024multivariate}
X.~Sun, H.~Liu, Multivariate short-term wind speed prediction based on pso-vmd-se-iceemdan two-stage decomposition and att-s2s, Energy 305 (2024) 132228.
\newblock \href {https://doi.org/10.1016/j.energy.2024.132228} {\path{doi:10.1016/j.energy.2024.132228}}.

\bibitem{al-musaylh2018two}
M.~S. AL-Musaylh, R.~C. Deo, Y.~Li, J.~F. Adamowski, Two-phase particle swarm optimized-support vector regression hybrid model integrated with improved empirical mode decomposition with adaptive noise for multiple-horizon electricity demand forecasting, Applied Energy 217 (2018) 422--439.
\newblock \href {https://doi.org/10.1016/j.apenergy.2018.02.140} {\path{doi:10.1016/j.apenergy.2018.02.140}}.

\bibitem{xu2024enhanced}
D.~Xu, J.~Yin, An enhanced hybrid scheme for ship roll prediction using support vector regression and tvf-emd, Ocean Engineering 307 (2024) 117951.
\newblock \href {https://doi.org/10.1016/j.oceaneng.2024.117951} {\path{doi:10.1016/j.oceaneng.2024.117951}}.

\bibitem{cao2023hybrid}
Z.~Cao, J.~Wang, L.~Yin, D.~Wei, Y.~Xiao, A hybrid electricity load prediction system based on weighted fuzzy time series and multi-objective differential evolution, Applied Soft Computing 149 (2023) 111007.
\newblock \href {https://doi.org/10.1016/j.asoc.2023.111007} {\path{doi:10.1016/j.asoc.2023.111007}}.

\bibitem{Chenoilprice}
Y.~Chen, Z.~Tian, A crude oil price forecasting framework based on constraint guarantee and pareto fronts shrinking strategy, Applied Soft Computing 174 (2025) 112996.
\newblock \href {https://doi.org/10.1016/j.asoc.2025.112996} {\path{doi:10.1016/j.asoc.2025.112996}}.

\bibitem{tian2025new}
Z.~Tian, M.~Gai, A new paradigm based on wasserstein generative adversarial network and time-series graph for integrated energy system forecasting, Energy Conversion and Management 326 (2025) 119484.
\newblock \href {https://doi.org/10.1016/j.enconman.2025.119484} {\path{doi:10.1016/j.enconman.2025.119484}}.

\bibitem{LuFrozen}
K.~Lu, A.~Grover, P.~Abbeel, I.~Mordatch, Frozen pretrained transformers as universal computation engines, Proceedings of the AAAI Conference on Artificial Intelligence 36 (2022) 7628--7636.
\newblock \href {https://doi.org/10.1609/aaai.v36i7.20729} {\path{doi:10.1609/aaai.v36i7.20729}}.

\bibitem{goswami2024parameter}
J.~Goswami, K.~K. Prajapati, A.~Saha, A.~K. Saha, Parameter-efficient fine-tuning large language model approach for hospital discharge paper summarization, Applied Soft Computing 157 (2024) 111531.
\newblock \href {https://doi.org/10.1016/j.asoc.2024.111531} {\path{doi:10.1016/j.asoc.2024.111531}}.

\bibitem{Zhou2023}
T.~Zhou, P.~Niu, X.~Wang, L.~Sun, R.~Jin, One fits all: Power general time series analysis by pretrained lm, Vol.~36, 2023, cited by: 165.

\bibitem{CriticalDayMatters}
W.~Liu, Z.~Tian, J.~Cui, C.~Wu, Load forecasting with deep learning: Critical day matters, in: 2024 IEEE Power \& Energy Society General Meeting (PESGM), 2024, pp. 1--5.
\newblock \href {https://doi.org/10.1109/PESGM51994.2024.10688616} {\path{doi:10.1109/PESGM51994.2024.10688616}}.

\bibitem{TONG2020102082}
D.~Tong, Y.~Zhang, I.~MacLachlan, G.~Li, Migrant housing choices from a social capital perspective: The case of shenzhen, china, Habitat International 96 (2020) 102082.
\newblock \href {https://doi.org/https://doi.org/10.1016/j.habitatint.2019.102082} {\path{doi:https://doi.org/10.1016/j.habitatint.2019.102082}}.

\end{thebibliography}

\end{document}